\documentclass[twocolumn]{aastex61}
\shorttitle{Comparing M Dwarf Models}
\shortauthors{Dieterich et al.}

\begin{document}

\title{The Solar Neighborhood XLVII: Comparing M Dwarf Models with Hubble Space Telescope Dynamical Masses and Spectroscopy}\footnote{Based on
    observations made with the NASA/ESA Hubble Space Telescope, obtained at the Space Telescope Science Institute,
    which is operated by the Association of Universities for Research in Astronomy, Inc. under NASA contract NAS5-26555. These
observations are associated with program no. 12938.}

\correspondingauthor{Serge B. Dieterich}
\email{sbdieterich@gmail.com}

\author{Serge B. Dieterich}
\altaffiliation{NSF Astronomy and Astrophysics Postdoctoral Fellow}
\affil{Department of Terrestrial Magnetism, Carnegie Institution \\
5241 Broad Branch Road, NW \\
Washington, DC 20015-1305, USA}
\altaffiliation{RECONS Institute \\
  Chambersburg, PA, USA}
\altaffiliation{Current address: Space Telescope Science Institute, Baltimore, MD, USA}

\author{Andrew Simler}
\affil{Reed College, Portland, OR, USA}

\author{Todd J. Henry}
\affil{RECONS Institute \\
 Chambersburg, PA, USA}

\author{Wei-Chun Jao}
\affil{Georgia State University \\
  Atlanta, GA, USA}

\begin{abstract}
  We use {\it HST/STIS} optical spectroscopy of ten M dwarfs in five
  closely separated binary systems to test models of M dwarf structure
  and evolution. Individual dynamical masses ranging from
  0.083\,M$_{\odot}$ to 0.405\,M$_{\odot}$ for all stars are known
  from previous work. We first derive temperature, radius, luminosity,
  surface gravity, and metallicity by fitting the BT-Settl atmospheric
  models. We verify that our methodology agrees with empirical results
  from long baseline optical interferometry for stars of similar
  spectral types.  We then test whether or not evolutionary models can
  predict those quantities given the stars' known dynamical masses and
  the conditions of coevality and equal metallicity within each binary
  system.  We apply this test to five different evolutionary model
  sets: the Dartmouth models, the MESA/MIST models, the models of
  \citet{Baraffeetal2015}, the PARSEC models, and the YaPSI models. We
  find marginal agreement between evolutionary model predictions and
  observations, with few cases where the models respect the condition
  of coevality in a self-consistent manner.  We discuss the pros and
  cons of each family of models and compare their predictive power.
 \end{abstract}

\section{Introduction} \label{sec:intro}

One of the principal goals of science is to explain the inner workings
of nature through the development of theoretical models that can then
be tested against the results of experiment and observation.
Concerning solar and higher mass stars, the overall theory of stellar
structure and evolution, as developed through most of the 20$^{th}$
century, is a triumph. From this theoretical framework we are able to
explain a star's locus in the Hertzsprung-Russel diagram as well as
its evolution before, during, and after the main sequence. The theory
of stellar structure has achieved such accuracy that we now often rely
on models of stellar evolution to calibrate observations and not the
other way around, such as when determining the ages of clusters by
using their main sequence turn-off points. The success of the theory
is due in part to the simplicity of the physics involved. As
extremely hot objects, stars are high entropy systems. This paradigm
deteriorates as one approaches lower stellar masses and cooler temperatures,
where effects such as convection and molecular impediments to
radiative transfer become more relevant. That is the realm of the M
dwarfs, with masses of 0.08\,M$_{\odot}$ to 0.62\,M$_{\odot}$ \citep{Benedictetal2016},
where several aspects of stellar theory are still not
precisely settled.

Early theoretical attempts at modeling M dwarf interiors include
\citet{Osterbrock1953} and \citet{Limber1958}, which were the first
treatments to include convection.  These early works relied on gray
model atmospheres that were poor approximations for radiative transfer
boundary conditions. Further advancements then followed with more thorough
spectroscopic characterizations \citep{Boeshaar1976} and the
formulation of non-gray model atmospheres \citep{Mould1976}. More
recently, the advent of large infrared photometric surveys such as
2MASS \citep{Skrutskieetal2006}, spectroscopic surveys like SDSS
\citep{Yorketal2000} and of M dwarf mass-luminosity relations
\citep{HenryandMcCarthy1993, Henryetal1999, Delfosseetal2000,
  Benedictetal2016} have provided a wealth of data, making the field
ripe for substantial advancements.  The development of sophisticated
model atmospheres for cool stars, amongst them the ones computed with
the PHOENIX code \citep{Hauschildtetal1997}, has greatly ameliorated
the treatment of the outer boundary conditions and allows for
the derivation of the fundamental parameters of effective temperature,
metallicity, and surface gravity solely from spectroscopic data. These
advances led to the generation of several families of low mass
evolutionary models (Sections \ref{subsec:evolutionary} and
\ref{subsec:approaches}) that are now widely used to estimate M dwarf
parameters and to construct synthetic stellar populations.

In this paper we use spatially resolved spectroscopic observations of
ten M dwarfs in five binary systems, all with precise dynamical
masses, to test the predictions of five models of stellar structure
and evolution: the Dartmouth models \citep{Dotteretal2008}, the MIST
models \citep{Choietal2016,Dotteretal2016}, the models of
\citet{Baraffeetal2015}, the PARSEC models \citep{Bressanetal2012},
and the YaPSI models \citep{Spadaetal2017}.  We begin by assessing the
quality of the BT-Settl model atmospheres \citep{Allardetal2012,
  Allardetal2013} and upon verifying the validity of their fits to
observed spectra, use them to infer effective temperature,
metallicity, and surface gravity. We then test whether or not
evolutionary models can replicate those values given the known
dynamical mass of each component and the requirements of coevality and
equal metallicity for a given binary system. The paper is organized as
follows.  We describe our observations in Section \ref{sec:obs} and
data reduction in Section \ref{sec:reduction}.  We evaluate the
quality of the BT-Settl model atmospheres based on comparison to
results obtained with long baseline optical interferometry and derive
atmospheric fundamental parameters in Section
\ref{subsec:atmospheric}.  We test the evolutionary models in Section
\ref{subsec:evolutionary}.  We discuss the noteworthy GJ 22 system,
radius inflation, and the effect of small changes in mass and
metallicity in Section \ref{sec:discussion}.  We discuss our
conclusions and summarize our results in Sections
\ref{sec:conclusions} and \ref{sec:summary}.

\section{Observations} \label{sec:obs}
We obtained spatially resolved intermediate resolution ($R \sim 10,000$)
red optical spectroscopy for the components of five binary systems
using the Space Telescope Imaging Spectrograph ({\it STIS}) on the  {\it Hubble
Space Telescope} through program 12938.

Table \ref{tab:astrometric} lists the astrometric properties of the
five star systems observed. All systems were astrometrically
characterized in \citet{Benedictetal2016}. That work relied primarily
on observations taken with the Fine Guidance Sensors ({\it FGS}) on
the {\it Hubble Space Telescope}.  Because {\it FGS} measures
displacements relative to distant ``fixed'' stars it can map the
motion of each component of a binary system relative to the sidereal
frame, thus allowing for the determination of individual component
masses. We selected M dwarf systems to cover a broad range of masses
and with separations suitable for spatially resolved spectroscopy with
{\it HST/STIS} based on the preliminary unpublished results of
\citet{Benedictetal2016}.  In this work we use the trigonometric
parallaxes derived in \citet{Benedictetal2016} rather than the more
recent {\it Gaia} DR2 parallaxes \citep{GaiaDR22018} because the
latter use an astrometric model suitable for a single point source
whereas \citet{Benedictetal2016} solve parallax and orbital motion
simultaneously.  We make an exception and use the Gaia DR2 parallax in
the case of the GJ 1245 system.  The system is a hierarchical binary
with the B component widely separated from the AC component and
clearly resolved.  Whereas \citet{Benedictetal2016} publish a parallax
of 219.9$\pm$0.5\,$mas$ for GJ 1245 AC Gaia DR2 provides
213.13$\pm$0.6\,$mas$ for the AC component and 214.52$\pm$0.08 $mas$
for the B component. Assuming a negligible difference in distance
between the B and the AC components, the agreement in their Gaia
parallaxes indicates that the parallax of \citet{Benedictetal2016} for
the AC component may be off by as much as $\sim$7\,$mas$.  Because we
expect some error to be introduced in the Gaia DR2 parallax of the AC
component due to its unresolved multiplicity, we adopt the Gaia DR2
parallax for GJ 1245 B as the best estimate of the true parallax of
the unresolved AC component.  We also notice a large
discrepancy for the GJ 469 system, for which Gaia DR2 and
\citet{Benedictetal2016} publish parallaxes of 68.62$\pm$0.89\,$mas$
and 76.4$\pm$0.5\,$mas$, respectively. However, in the case of GJ 469
we have no reason to doubt the result from \citet{Benedictetal2016}
and Gaia DR2 parallaxes with uncertainties larger than $\sim$0.4 $mas$
are known to be suspect \citep{Vrijmoetetal2020}.  The
parallaxes for the other three systems agree to at least 3 percent in
distance.

In order to observe both components simultaneously it was necessary to
align the {\it STIS} long slit along the system's position angle, which
required rotating the {\it HST} spacecraft.  We calculated tables of
position angles for each system and matched them to HST's roll angle
time windows, which are determined by the need to keep the solar
panels exposed to sunlight. The observations were taken using the
G750M grating and the 0\farcs2 wide long slit.  The observations covered
the spectral range from 6,483\,\AA$~$ to 10,126\,\AA$~$ using seven grating
tilts. A contemporaneous W lamp flat was obtained before each exposure
to correct the fringing present in the STIS ccd at wavelengths greater
than 7,000\,\AA.  Individual exposure times ranged from 252\,s to
435\,s; however, two HST orbits were required per system to accommodate
the large overheads associated with changing grating tilts.

\section{Data Reduction} \label{sec:reduction}
The strategy used for data reduction depended on whether or not the
signal from both components was significantly blended in the spatial
direction. The components of G250-29, GJ 22, and  GJ 1245  were
sufficiently separated ($\gtrsim$ 0\farcs5) so that a saddle point with flux comparable to
the sky background could be identified (Table \ref{tab:astrometric}).
For these systems we used symmetry arguments to perform the sky
subtraction while also subtracting any residual flux from the opposite
component. The signal in the apertures for each component was then
reduced using the standard {\it calstis} pipeline provided by the Space
Telescope Science
Institute\footnote{http://www.stsci.edu/hst/instrumentation/stis/data-analysis-and-software-tools}
\citep{STISDataBook}. STIS is periodically flux calibrated with known
flux standards. The stability of the space environment precludes the need
for observing flux standards in close proximity to science observations.
{\it Calstis}
automatically performs flat fielding, bias and dark subtraction,
spectral extraction, wavelength calibration, flux calibration, and
1-dimensional rectification. 

The spectra for the components of GJ 469 GJ and 1081  were separated by
$\sim$0\farcs15,
less than three pixels in the spatial direction. The {\it calstis} pipeline
is meant for resolved point sources, and was therefore inadequate for
the deblending of these spectra.  For these sources we used a
subsampled synthetic {\it STIS} Point Spread Function (PSF) generated with
the Tiny Tim {\it HST} optical simulator\footnote{http://www.stsci.edu/software/tinytim/}
\citep{Kristetal2011} to replicate
the convoluted spectra.  Two synthetic PSFs subsampled by a factor of
ten were superimposed with an initial separation and flux ratio
estimated from the data. The separation and flux ratio were then
varied until the best correlation was obtained between the model and
the observed spectra. To account for the wavelength dependence of the
PSF we produced a new PSF for each 100\,\AA$~$ segment of the spectra.
The STIS ccd is
subject to considerable pixel crosstalk that is not modeled by Tiny
Tim when PSFs are subsampled. We approximated a crosstalk correction
by applying the known {\it STIS} crosstalk kernel to the best results of the
synthetic PSF scaling and then repeating the process, but adding the
crosstalk flux from the first iteration to this second iteration. As
expected, the crosstalk had the effect of slightly smoothing the
resulting spectra.

The {\it STIS} ccd exhibits considerable fringing starting at
wavelengths longer than 7,000\,\AA, and reaches an
amplitude of about 30 percent at wavelengths redward of
9,000\,\AA. The fringing can be largely subtracted using
contemporaneous flats taken with the on-board W calibration lamp. The
standard de-fringing procedure for point sources
\citep{Goudfrooijetal1998a,Goudfrooijetal1998b} assumes a smooth
spectrum with sharp absorption or emission lines that can be used to
optimally position the fringe pattern in the spectral direction. This
method was not suitable for M dwarfs due to the complex nature of
their spectra, where line blanketing precludes the continuum. We
devised a solution by extracting a fringe spectrum using a three pixel
aperture centered at the peak of the science spectrum and then scaling
the fringe spectrum until we obtained the least correlation between
the science spectrum and the fringe spectrum.  While this procedure
largely eliminated fringing at wavelengths bluer than 9,000\,\AA, as
shown in Figures \ref{fig:spectra1} and \ref{fig:spectra2}, fringing remains an issue at the
reddest wavelengths. However, because no model spectrum is likely to be any better
or worse in replicating the fringe noise, this fringing does not
interfere with our goal of finding the best model match to each
observed spectra. The {\it STIS} team at STScI is currently developing a new
defringing package that should further reduce the fringing\footnote{STIS team,
  personal communication}. Users who would like a better fringe correction are
encouraged to download the data from the {\it HST} archive and re-reduce it once
better defringing tools are available.

Aside from the fringing, {\it HST} observations of these relatively bright
sources are
subject to very little sky background and other sources of noise. As
we discuss in Section \ref{subsec:atmospheric} a detailed line to line
comparison with models shows that the observed spectra are rich in
fine structure. From that we estimate a signal to noise of 30 to 60 for the spectra,
depending on the source brightness and the wavelength region.

\begin{deluxetable*}{cccccccccc}
\tabletypesize{\scriptsize}
\tablecaption{Astrometric Properties \tablenotemark{a} \label{tab:astrometric}}
\tablehead{ \colhead{System}        &
            \colhead{RA}            &
            \colhead{Dec}           &
            \colhead{Parallax}      &
            \colhead{Semi-major\tablenotemark{b}}    &
            \colhead{Period}        &
            \colhead{Primary}       &
            \colhead{Secondary}     &
            \colhead{Date}          &
            \colhead{Approx. Separation}     \\
            \colhead{  }          &
            \colhead{(2000)}      &
            \colhead{(2000)}      &
            \colhead{{\it mas}}   &
            \colhead{Axis ({\it mas})}   &
            \colhead{Days}              &
            \colhead{Mass ($M_{\odot}$)} &
            \colhead{Mass ($M_{\odot}$)} &
            \colhead{Obs.}      &
            \colhead{{\it mas}}   }
\startdata
GJ 22 AC                      &  00 32 29.5 & +67 14 03.6 &  99.2$\pm$0.6 & 510.6$\pm$0.7 & 5694.2$\pm$14.9  & 0.405$\pm$0.008 & 0.157$\pm$0.003 & 2013-01-14 & 491 \\
GJ 1081 AB                    &  05 33 19.1 & +44 48 57.7 &  65.2$\pm$0.4 & 271.2$\pm$2.7 & 4066.1$\pm$27.5  & 0.325$\pm$0.010 & 0.205$\pm$0.007 & 2012-10-02 & 152 \\
G 250-29 AB                   &  06 54 04.2 & +60 52 18.3 &  95.6$\pm$0.3 & 441.7$\pm$0.9 & 4946.3$\pm$2.2   & 0.350$\pm$0.005 & 0.187$\pm$0.004 & 2013-01-15 & 517 \\     
GJ 469 AB                     &  12 28 57.4 & +08 25 31.2 &  76.4$\pm$0.5 & 313.9$\pm$0.8 & 4223.0$\pm$2.9   & 0.332$\pm$0.007 & 0.188$\pm$0.004 & 2013-03-24 & 152 \\      
GJ 1245 AC\tablenotemark{c}   &  19 53 54.4 & +44 24 53.0 & 213.1$\pm$0.6 & 826.7$\pm$0.8 & 6147.0$\pm$17    & 0.120$\pm$0.001 & 0.081$\pm$0.001 & 2013-06-04 & 598 \\ 
\enddata
\tablenotetext{a}{Values from \citet{Benedictetal2016} except date of HST/STIS observation and separation from that observation. See note (c) about GJ 1245.}
\tablenotetext{b}{Semi-major axis of relative orbit of secondary around primary component.}
\tablenotetext{c}{Parallax from {\it Gaia DR2}. Dynamical masses were adjusted to reflect that parallax. See See
  Section \ref{sec:obs} for a discussion of the GJ 1245 system.}
\end{deluxetable*}

\section{Results} \label{sec:results}
Figures \ref{fig:spectra1} and \ref{fig:spectra2} show the normalized spectrum for
each star along with the best matching model spectrum and the fit residual (Section
\ref{subsec:atmospheric}). Table \ref{tab:temp} outlines the derived properties for the
ten stars in the five systems. 
We obtained spectral type for individual
components
using the spectral type templates of
\citet{Bochanskietal2007}\footnote{https://github.com/jbochanski/SDSS-templates}
and performing a full spectrum $\chi^2$ minimization. A clear match to
a template was found in all cases except GJ 1081 B (M4.5V) and GJ 469
A (M3.5V), where we interpolated between the two best matches to
obtain a fractional subclass.

We next discuss detailed fits to atmospheric and evolutionary models.
Here we draw a sharp distinction between the two types of models in
the sense that we at first do not consider the internal stellar
parameters that govern stellar luminosity and dictate its evolution.
In other words, we assume that atmospheric model fits can tell us much
about the star accurately predict a star's temperature, radius,
luminosity, metallicity, and surface gravity without making any
theoretical assumptions about interior physics.  We validate the
accuracy of the atmospheric model fits in Section
\ref{subsubsec:interferometry}, where we compare the results of our
model fitting methodology to known radii, temperatures, and
luminosities measured with long baseline optical interferometry for a
sample of 21 calibrator stars.  We then use the known masses, the
observed flux, the trigonometric parallax, and the Stefan-Boltzmann
law to derive radii, luminosities, and surface gravities.  At that
point we connect the discussion to the predictions of structure and
evolution models by discussing what internal conditions could be the
cause of these observed fundamental properties.

\subsection{Fitting Atmospheric Models} \label{subsec:atmospheric}

Atmospheric models are one of the cornerstones of our understanding of
stellar physics because they provide an extremely rich set of
predictions (i.e., a synthetic spectrum) that can then be readily
tested with observed data. Here we compare the data to the BT-Settl
family of models \citep{Allardetal2013,Allardetal2012}. BT-Settl is a
publicly available and widely used implementation of the PHOENIX model
atmosphere code \citep{Hauschildtetal1997} that covers the relevant
temperature range (3,500\,K to 2,600\,K), is based on modern estimates
of solar metallicites \citep{Caffauetal2011}, and incorporates a grain
sedimentation cloud model, which is necessary in modeling cool M dwarf
atmospheres. Temperature is modeled in increments of 100\,K.
Metallicity ([Fe/H]) can take the values -1.0, -0.5, 0.0, and at some
grid elements +0.5, and $Log\,g$ ranges from 2.0 to 5.5 in increments
of 0.5.

We take the model fitting approach described in
\citet{MannGaidosAndAnsdell2013}.  We first trimmed the model grid to
include temperatures from 2,000\,K to 3,900\,K with no restrictions on
metallicity or surface gravity, resulting in a total of 335 model
spectra. We trimmed wavelengths to include the range from 6,000\,\AA
to 10,200\,\AA and applied a Gaussian smoothing kernel to smooth the
model spectra to the same resolution as the data. Because the
differences between the model spectra and the data are driven partly
by systematic errors in modeling a $\chi^2$ fit is not appropriate. To
find best fits we instead minimize the $G_K$ statistic, described in  \citet{Cushingetal2008} and
\citet{MannGaidosAndAnsdell2013}:
\begin{equation}
G_K = \sum_{i = 1}^{n} \left(\frac{w_i(F_i - C_K
  F_{K,i})}{\sigma_i}\right)^2
\end{equation} 
where $F_i$ is the data flux in the $i^{th}$ wavelength bin, $F_{K,i}$
is the model flux, and $C_K$ is a normalization constant. We set $C_K$
so that the mean of $F_i$ and $F_K$ are the same. $w_i$ is the weight
of the $i^{th}$ data element and $\sigma_i$ is its uncertainty.

\begin{figure*}[h!]
   \includegraphics[scale=0.55]{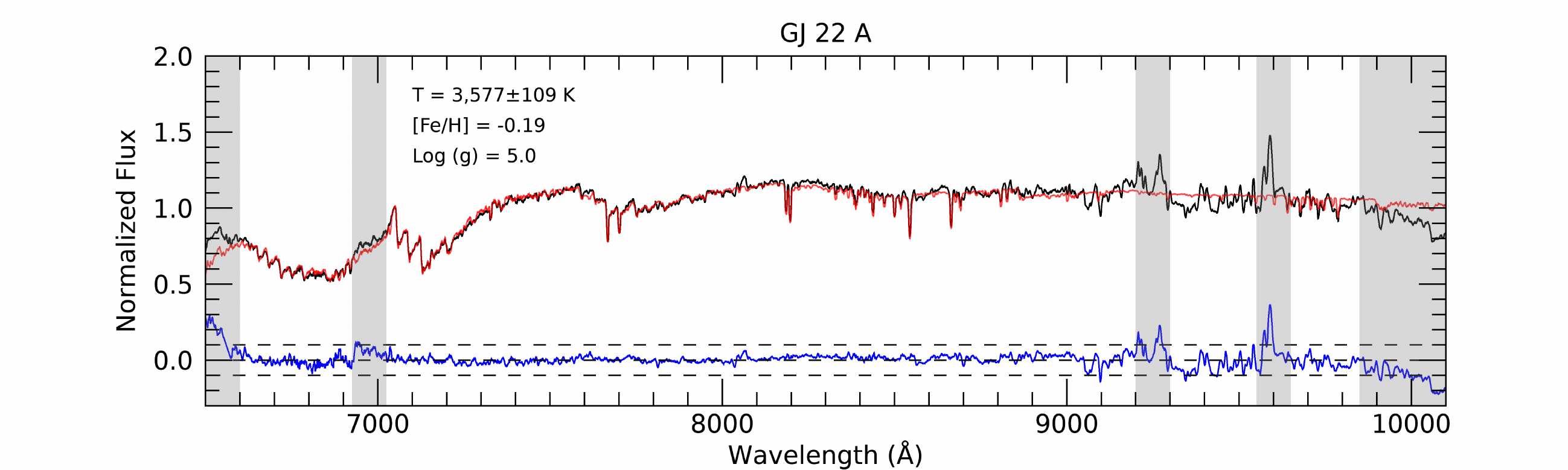}
   \includegraphics[scale=0.55]{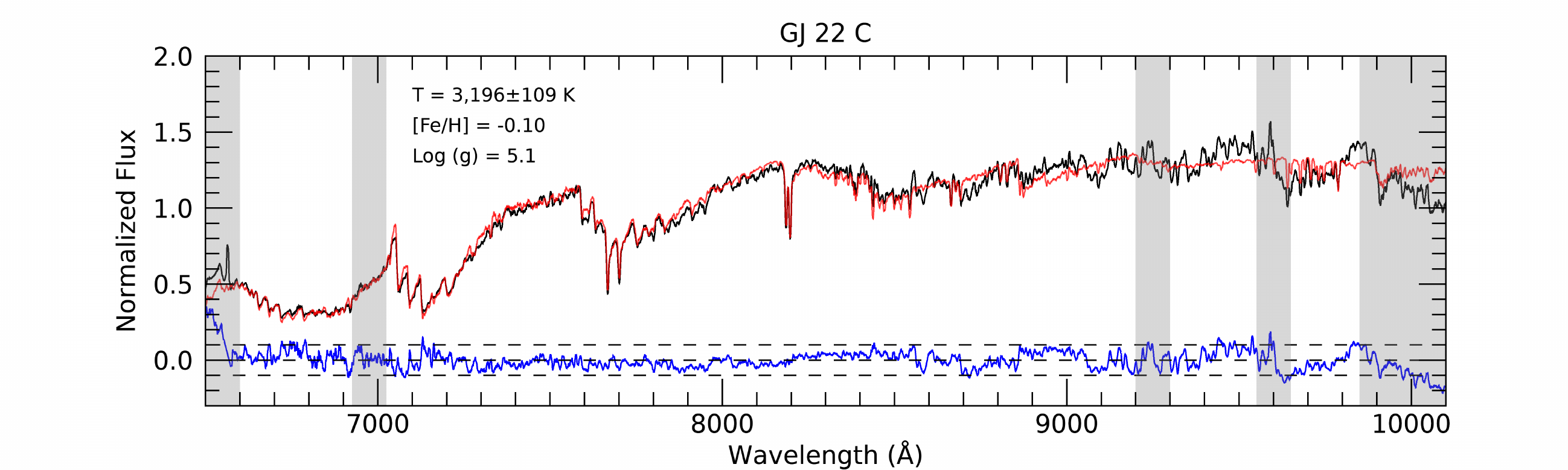}
   \includegraphics[scale=0.55]{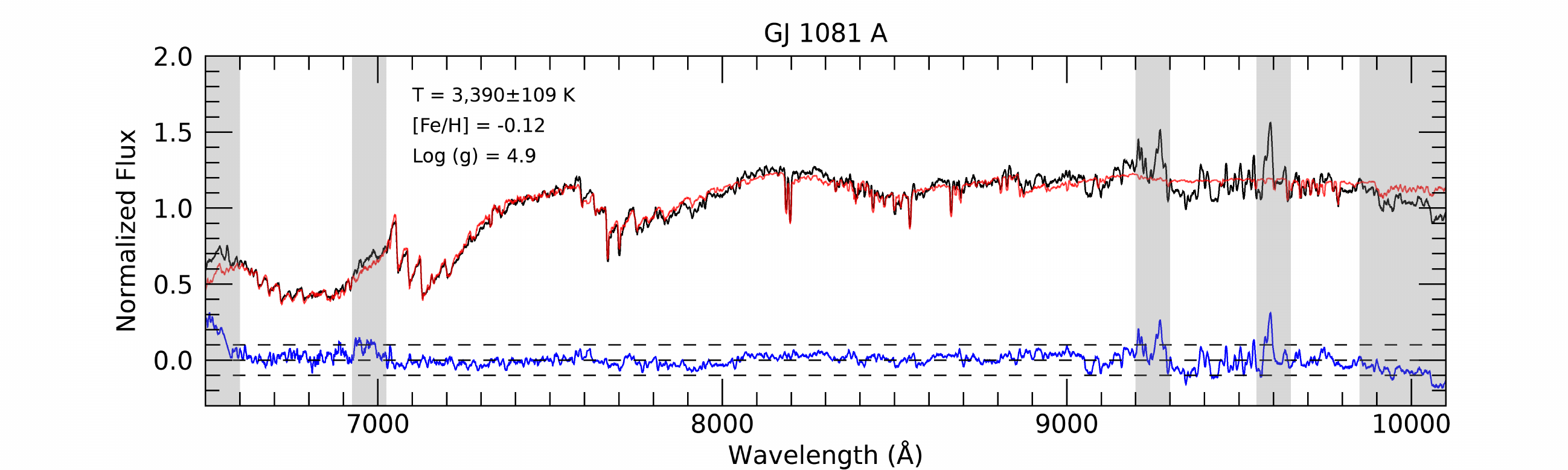}
   \includegraphics[scale=0.55]{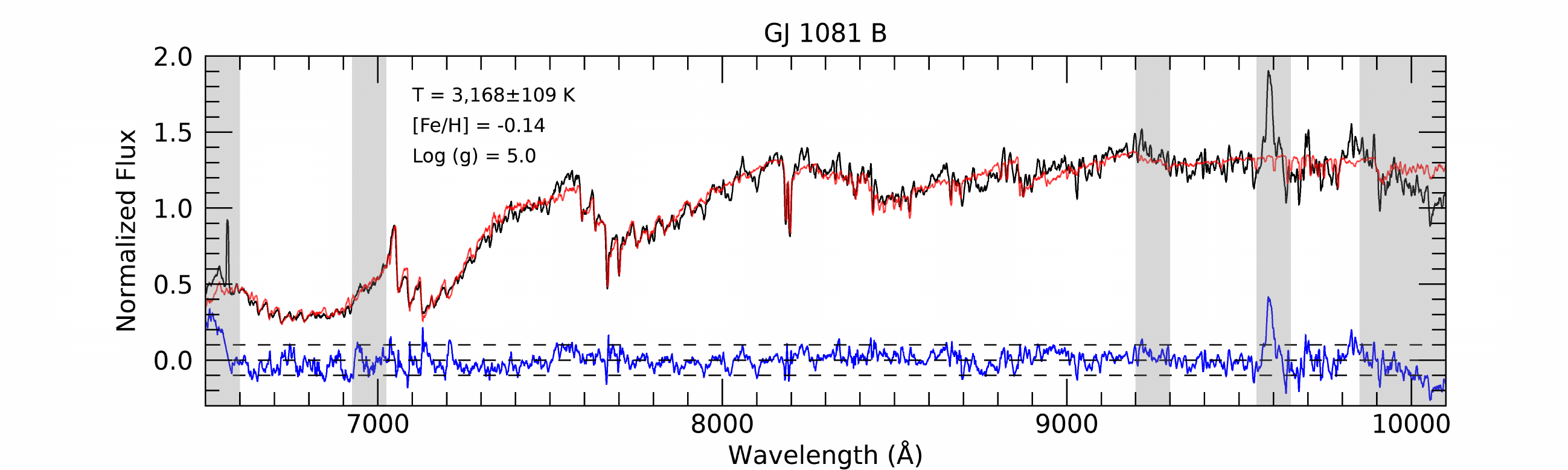}
   \includegraphics[scale=0.55]{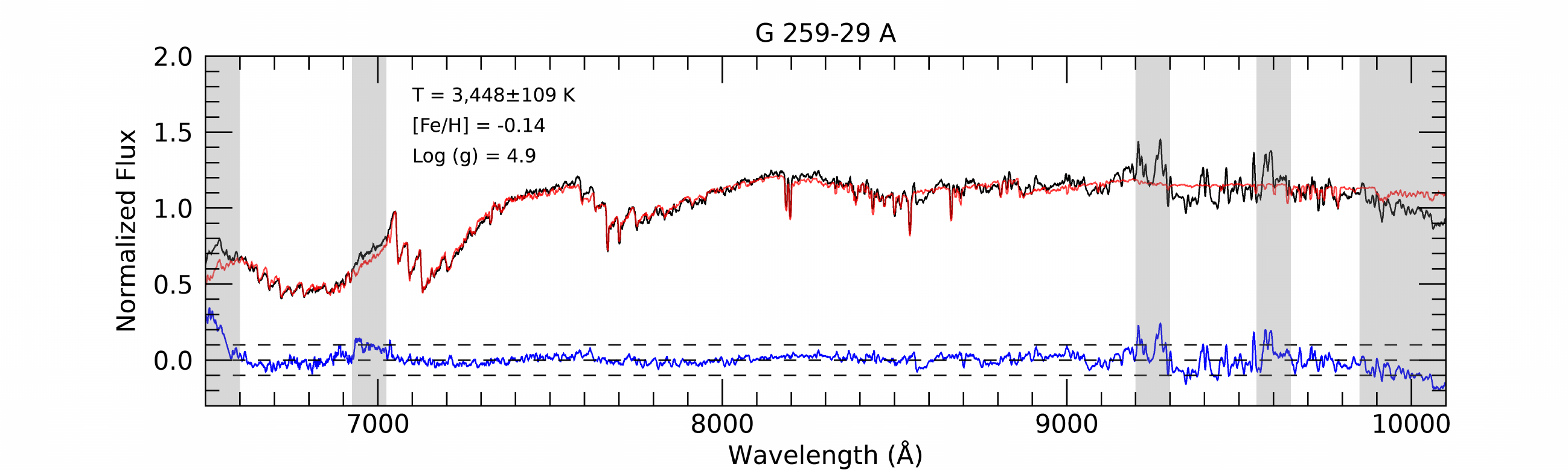}
   \caption{\scriptsize Normalized spectra are plotted in black with the best fitting
       combination model spectrum over-plotted in red. The shaded gray regions are wavelength
       ranges not used in the model fit. The fit residual is plotted in blue at the bottom,
       with the outer dashed lines showing the 10 percent residual mark. Both the observed and the
     model spectra were smoothed
     for clarity. High resolution unsmoothed images are available in the electronic version.
     Fringing is evident as step-like sharp features at wavelengths
     redder than 9,000\,\AA. \label{fig:spectra1}}
 \end{figure*}

\begin{figure*}[h!]
   \includegraphics[scale=0.55]{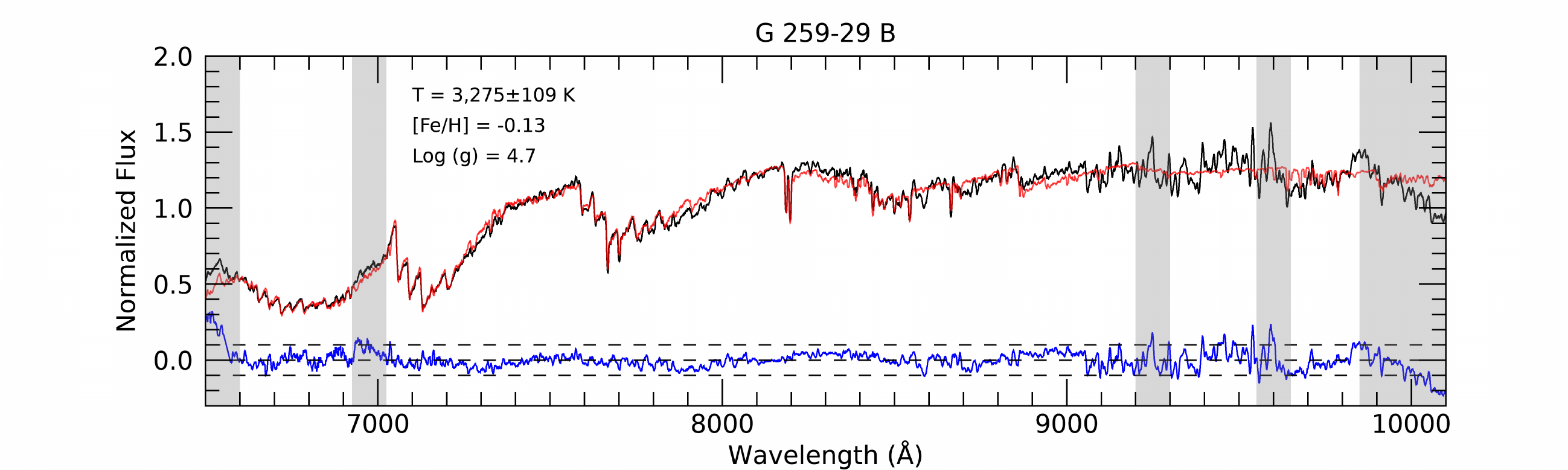}
   \includegraphics[scale=0.55]{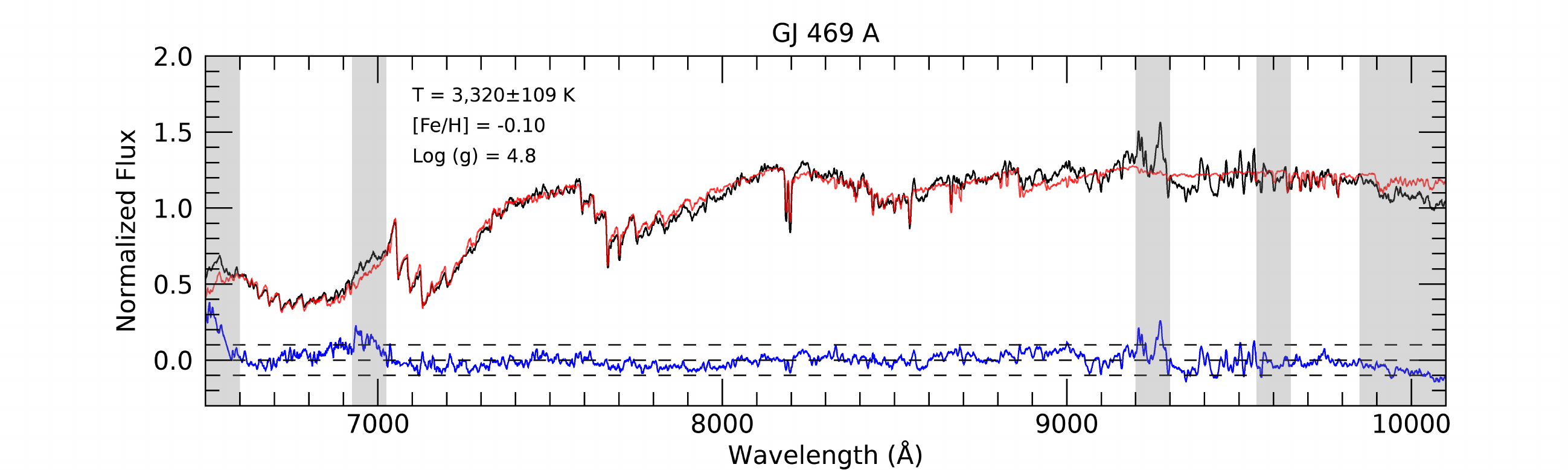}
   \includegraphics[scale=0.55]{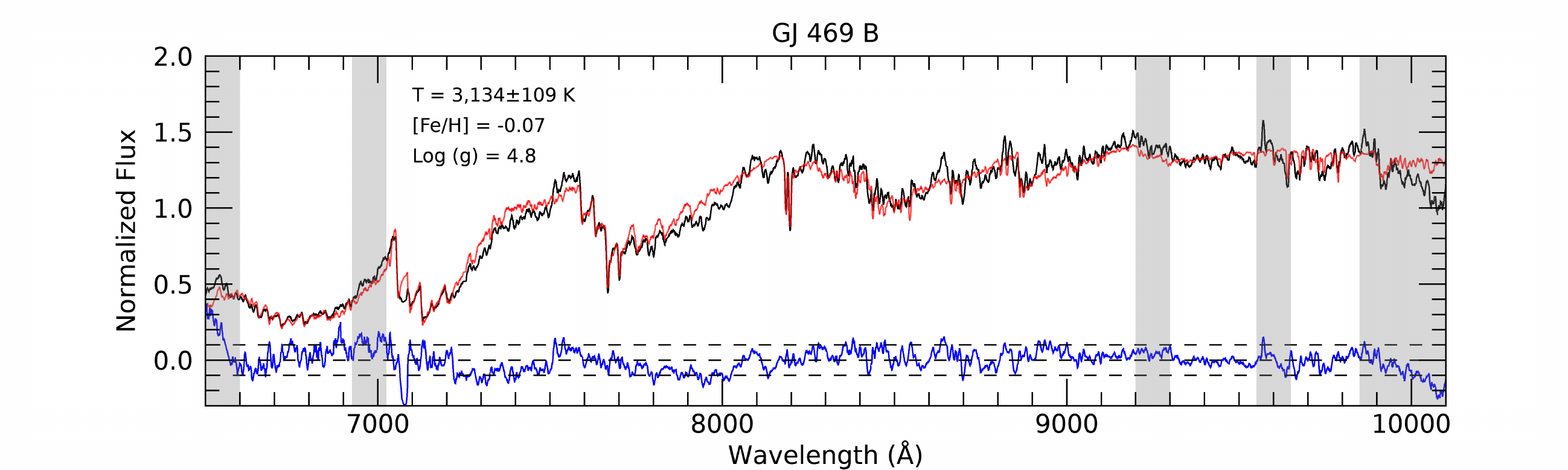}
   \includegraphics[scale=0.55]{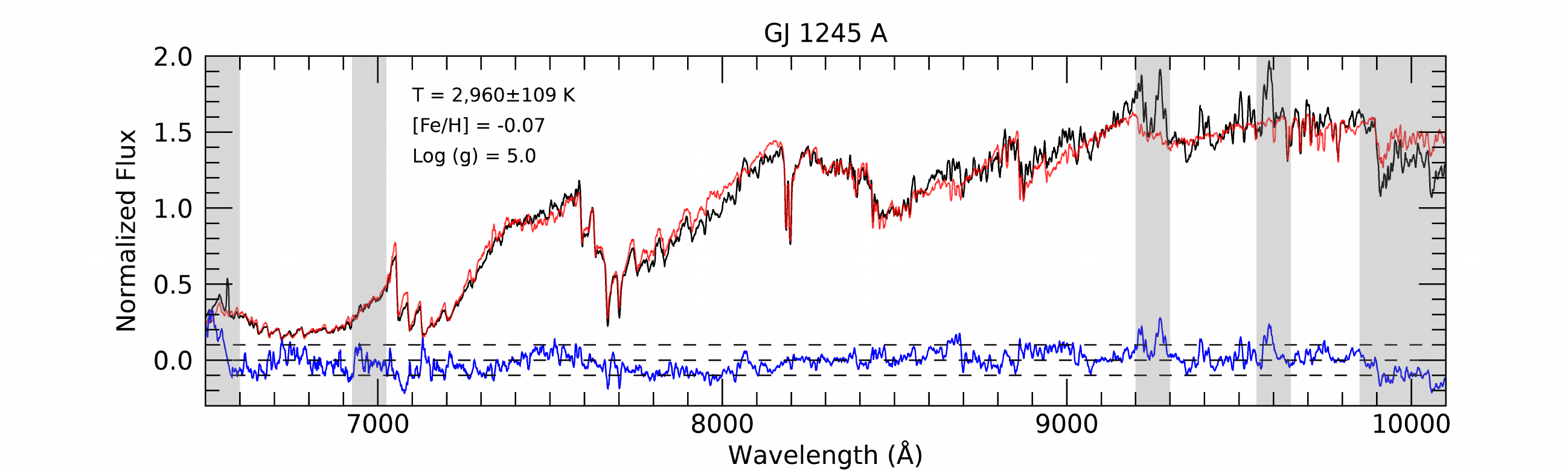}
   \includegraphics[scale=0.55]{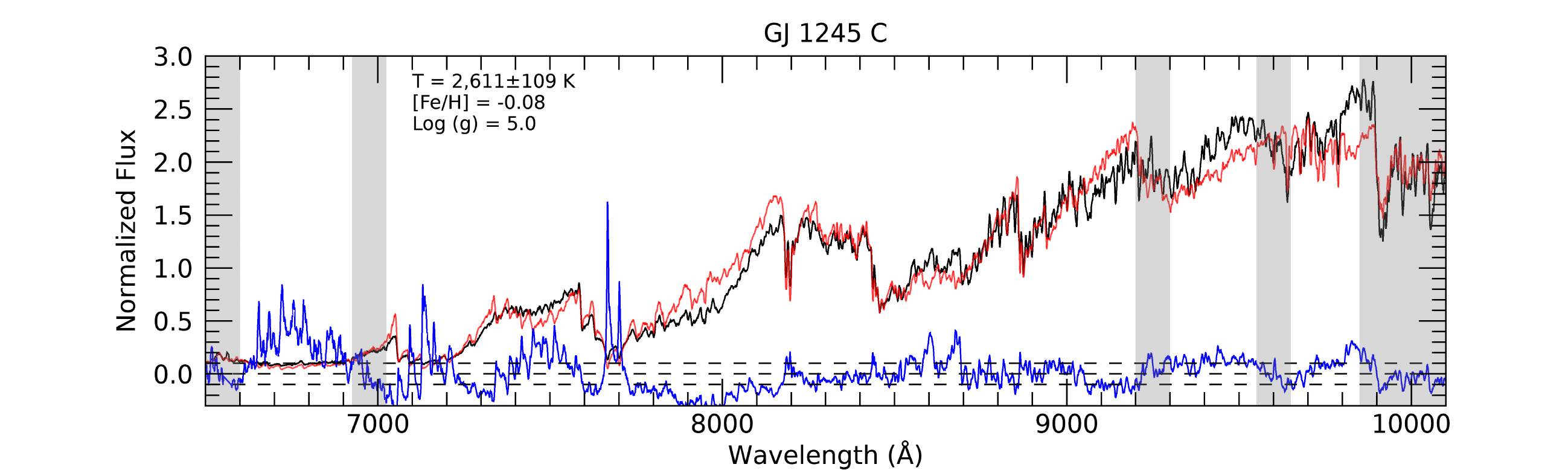}
   \caption{\scriptsize The continuation of Figure \ref{fig:spectra1}. The quality of the fits is
     degraded at lower temperatures, as is evident for GJ 1245 A and C.
     \label{fig:spectra2}}
 \end{figure*}

We first perform an initial fit where for each star we rank all 335
model spectra by minimizing $G_K$ with all weights $w_i$ set equal to
one. Because the signal-to-noise is high in all cases it does not
substantially alter the fit, and we set $\sigma$ corresponding to a
signal-to-noise of 50 for all elements.  We then select the top 20
best model fits and compute 10,000 random linear combinations, and
select the best linear combination via the $G_K$ minimization again
with all weights set to one. We then compute the residuals of the fit
to each of the ten stars and take the mean of all ten residuals. We
note regions where the mean residual is greater than 10 percent for
10\,\AA $~$ or more and re-iterate the process now setting $w_i = 0$ for
those regions.  The excluded wavelength regions are: 6,483\,\AA $~$to
6,600\,\AA, 6,925\,\AA $~$to 7,025\,\AA, 9,300\,\AA $~$to
9,400\,\AA, 9,550\,\AA $~$to 9,650\,\AA $~$and 9,850\,\AA $~$to
10,126\,\AA, with the third and fourth region due to strong fringing
in the data.  Figures \ref{fig:spectra1} and \ref{fig:spectra2} show
the resulting model fits superimposed on the the normalized spectra
and the corresponding residuals, with the traces smoothed for
clarity.  The online supplement to Figures \ref{fig:spectra1} and
\ref{fig:spectra2} shows full resolution spectra and model fits on a
flux calibrated scale. As described in Section  \ref{subsubsec:interferometry},
this fitting method produces a standard deviation of 109\,K in temperature
when compared to effective temperatures derived from long baseline optical
interferometry, and we adopt that as the uncertainty in the temperatures
we report in Table \ref{tab:temp}.

\begin{deluxetable*}{cccccccccccc}[h!]
\tabletypesize{\scriptsize}
\tablecaption{Derived Properties \label{tab:temp}}
\tablehead{ \colhead{Star}                                &
            \colhead{Mass \tablenotemark{a}}   &
            \colhead{$M_V$ \tablenotemark{b}}       &
            \colhead{$M_K$ \tablenotemark{b}}       &
            \colhead{Spectral}                       &
            \colhead{Temperature}  &
            \colhead{$Log$ $g$   }   &
            \colhead{[$Fe/H$]    }       &
            \colhead{Radius } &
            \colhead{Luminosity  }      &
            \colhead{$Log$ $g$ \tablenotemark{c}}  &
            \colhead{H$\alpha$}        \\
            \colhead{   }         &
            \colhead{($M_{\odot}$)} &
            \colhead{   }         &
            \colhead{   }         &
            \colhead{Type}         &
            \colhead{$K$}         &
            \colhead{Fit}         &
            \colhead{   }         &
            \colhead{($R_{\odot}$)} &
            \colhead{($Log(L/L_{\odot})$)} &
            \colhead{Calculated}    &
            \colhead{EW}    }
\startdata
GJ 22 A     &  0.405$\pm$0.008  & 10.32$\pm$0.03  & 6.19$\pm$0.02  &  M2V      &  3,577          &  5.0       &   -0.19    &   0.376$\pm$0.018 &  -1.68$\pm$0.09 &  4.9      &  0.40  \\
GJ 22 C     &  0.157$\pm$0.003  & 13.40$\pm$0.10  & 8.12$\pm$0.04  &  M4V      &  3,196          &  5.1       &   -0.10    &   0.179$\pm$0.009 &  -2.52$\pm$0.10 &  5.1      & -2.12  \\
GJ 1081 A   &  0.325$\pm$0.010  & 11.49$\pm$0.04  & 6.79$\pm$0.04  &  M3V      &  3,390          &  4.9       &   -0.12    &   0.343$\pm$0.015 &  -1.85$\pm$0.09 &  4.9      & -0.73  \\
GJ 1081 B   &  0.205$\pm$0.007  & 13.16$\pm$0.09  & 7.75$\pm$0.04  &  M4.5V    &  3,168          &  5.0       &   -0.14    &   0.237$\pm$0.011 &  -2.29$\pm$0.10 &  5.0      & -4.37  \\
G 250-29 A  &  0.350$\pm$0.005  & 11.07$\pm$0.03  & 6.61$\pm$0.03  &  M4V      &  3,448          &  4.7       &   -0.14    &   0.355$\pm$0.017 &  -1.79$\pm$0.10 &  4.9      &  0.32  \\
G 250-29 B  &  0.187$\pm$0.004  & 12.68$\pm$0.07  & 7.64$\pm$0.05  &  M3V      &  3,279          &  4.7       &   -0.11    &   0.231$\pm$0.011 &  -2.25$\pm$0.10 &  5.0      &  0.33  \\
GJ 469 A    &  0.332$\pm$0.007  & 11.69$\pm$0.03  & 6.74$\pm$0.04  &  M3.5V    &  3,320          &  4.8       &   -0.10    &   0.329$\pm$0.016 &  -1.93$\pm$0.10 &  4.9      &  0.27  \\
GJ 469 B    &  0.188$\pm$0.004  & 13.28$\pm$0.05  & 7.75$\pm$0.04  &  M5V      &  3,134          &  4.8       &   -0.07    &   0.266$\pm$0.011 &  -2.35$\pm$0.10 &  5.0      &  0.00  \\ 
GJ 1245 A   &  0.120$\pm$0.001  & 15.12$\pm$0.03  & 8.85$\pm$0.02  &  M6V      &  2,927          &  4.9       &   -0.07    &   0.146$\pm$0.007 &  -2.85$\pm$0.11 &  5.2      & -2.96  \\
GJ 1245 C   &  0.081$\pm$0.001  & 18.41$\pm$0.06  & 9.91$\pm$0.02  &  M8V      &  2,611          &  5.0       &   -0.08    &   0.087$\pm$0.004 &  -3.50$\pm$0.12 &  5.5      & -2.93  \\
\enddata
\tablenotetext{a}{All masses except for  GJ 1245 A and C are from \citet{Benedictetal2016}. The masses for the GJ
  1245 system have been corrected to reflect the more accurate {\it Gaia DR2} parallax. See Section \ref{sec:obs} for
  a discussion of the GJ 1245 system.}
\tablenotetext{b}{From \citet{Benedictetal2016} and references therein.}
\tablenotetext{c}{Calculated based on inferred radius and dynamical mass.}
  \end{deluxetable*}
 
  We calculated stellar radii in Table \ref{tab:temp} by scaling the
 model flux at the stellar surface and the observed flux via the
 geometric scaling relation $R_{\star}^2 = d^2 (F_{\oplus}/F_{\star})$
 where $R_{\star}$ is the stellar radius, $d$ is the trigonometric
 parallax distance to the star (Table \ref{tab:astrometric}), and
 $F_{\oplus}$ and $F_{\star}$ are the observed flux and the model flux
 at the stellar surface, respectively. We then calculated luminosities
 using the Stefan-Boltzmann law and surface gravities based on radii
 and masses. The latter serve as checks on the surface gravities
 assumed in the model spectra. We derive the uncertainties in
   radius in Section \ref{subsubsec:interferometry} and propagate the
   uncertainties in temperature and radius to obtain the uncertainties
   in luminosity.

\subsubsection{Validating the Atmospheric Model Fits with Long Baseline Optical Interferometry} \label{subsubsec:interferometry}
To prove the adequacy of our atmospheric model derived quantities
(Table \ref{tab:temp}) we again follow the procedure of
\citet{MannGaidosAndAnsdell2013}. For a calibrator sample of 21 stars
we compare effective temperatures and radii against the same
quantities derived based on angular diameters directly measured with
the CHARA Array long baseline optical
inteferometer\footnote{http://www.chara.gsu.edu}
\citep{Boyajianetal2012}.  Once angular diameters are measured via
interferometry stellar radii are trivially obtained given the well
known trigonometric parallaxes to these bright nearby stars. Effective
temperatures can be calculated via the Stefan-Boltzmann law if the
bolometric luminosity is known.  The latter can be well approximated
thanks to wide photometric and spectroscopic observations covering the
spectral energy distribution from the near ultraviolet to the mid
infrared. \citet{MannGaidosAndAnsdell2013} improved upon the
photometric treatment of \citet{Boyajianetal2012} to derive the
interferometric temperatures for the 21 stars listed in Table
\ref{tab:interferometry}.

\begin{figure*}[h!]
  \begin{center}
   \includegraphics[scale=0.45]{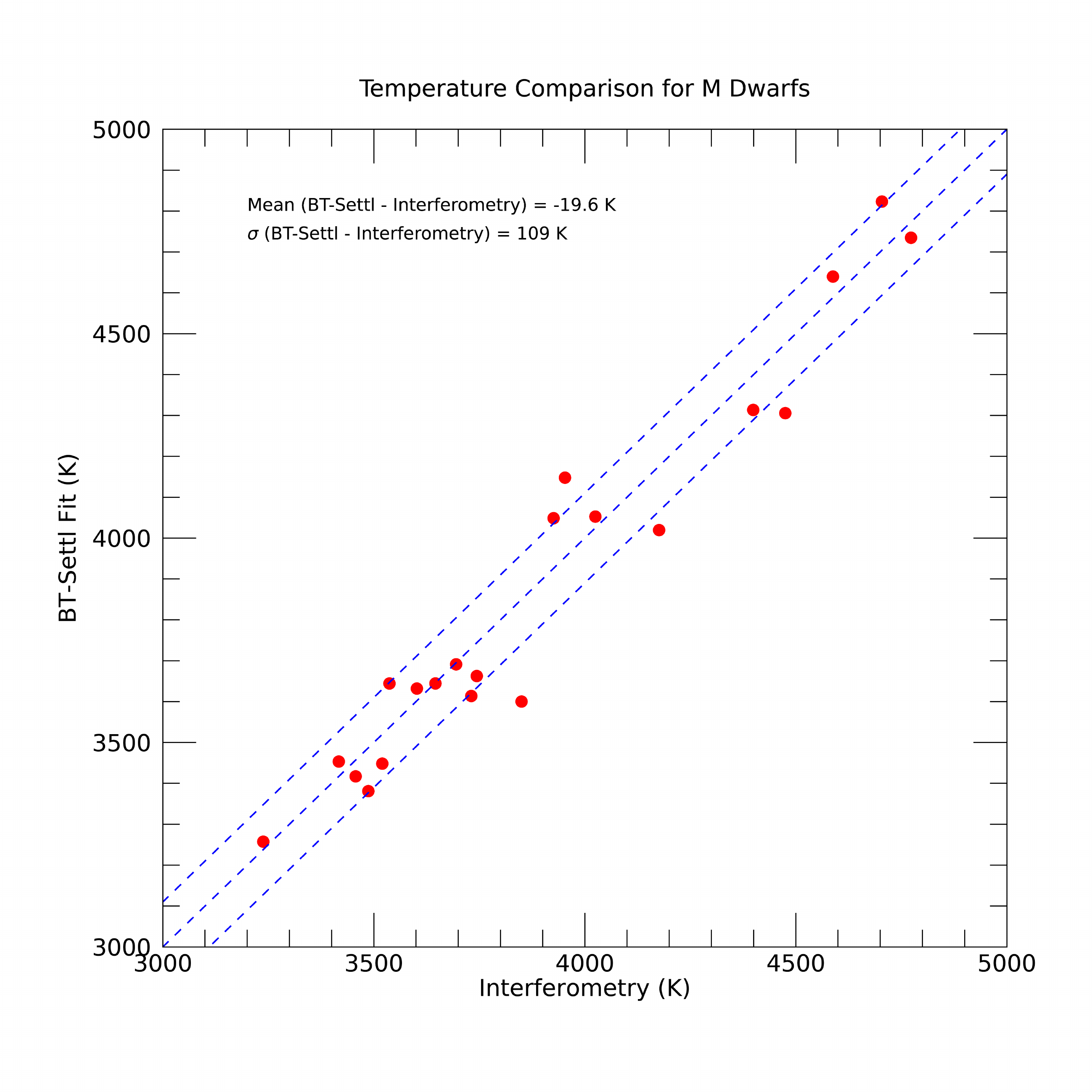}
   \caption{\scriptsize Comparison of effective
     temperatures obtained with long baseline optical interferometry and
     our atmospheric fitting technique for a calibrator sample of 21 M
     dwarfs. The blue dashed lines denote the standard deviation of 109 K.
     A Student's t test shows no systematic difference between
     the samples to 0.89 significance. \label{fig:tempinterferometry}}
\end{center}
\end{figure*}

\begin{figure*}[h!]
\begin{center}
  \includegraphics[scale=0.45]{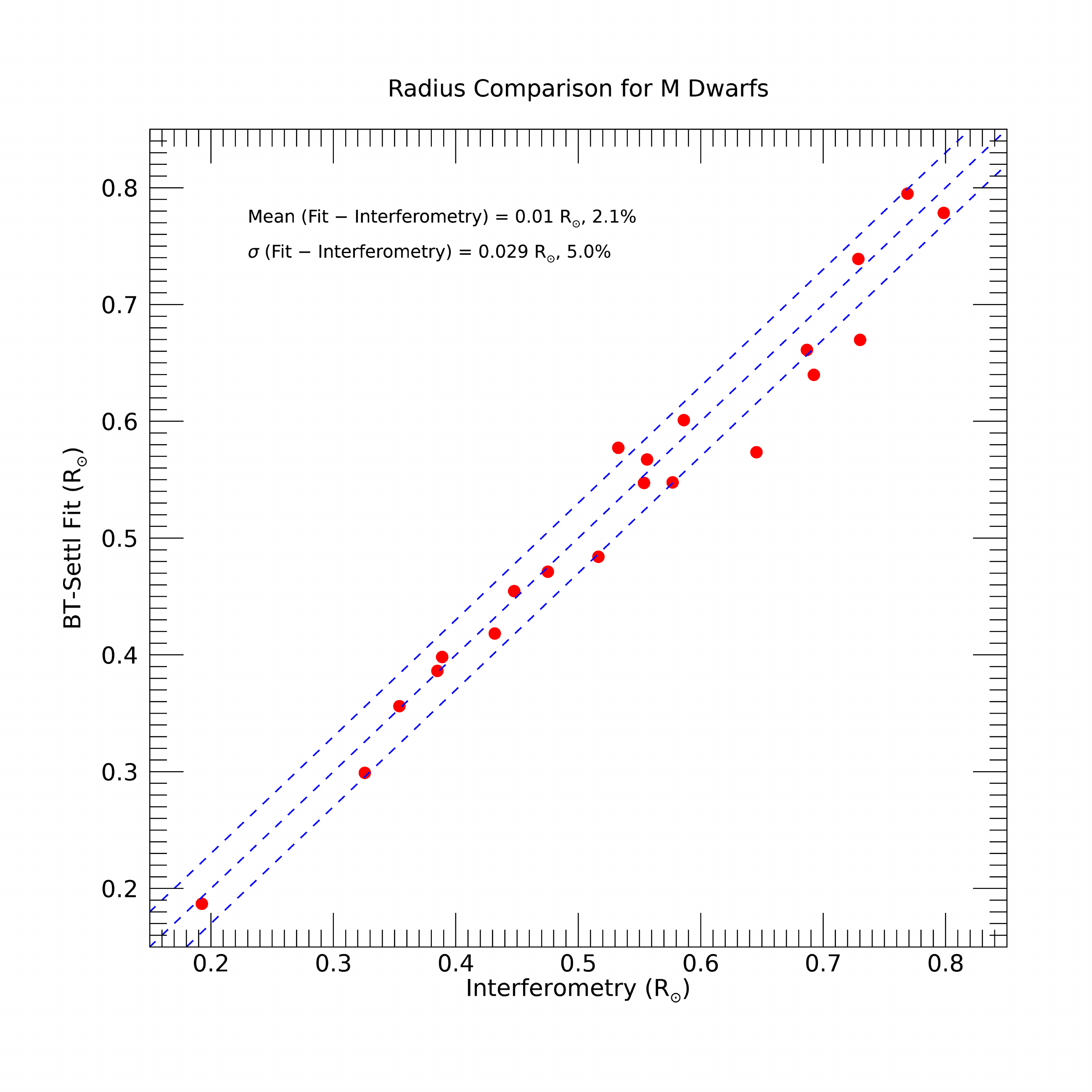}
   \caption{\scriptsize Same as
     Figure \ref{fig:tempinterferometry}, but for comparing radius.
     \label{fig:radiusinterferometry}}
\end{center}
\end{figure*}

Figures \ref{fig:tempinterferometry} and
\ref{fig:radiusinterferometry} show the comparison of effective
temperatures and radii, respectively, obtained with interferometry and
our atmospheric model fitting technique. The calibrator spectra from 
\citet{MannGaidosAndAnsdell2013} were kindly made available by Andrew Mann.

The procedure for fitting
this sample was the same as the one described in Section
\ref{subsec:atmospheric} except that we also excluded the wavelength
regions from 7,050\,\AA $~$to 7,150\,\AA $~$due to the effect of the
atmospheric oxygen A band. All other telluric regions appear to be
well accounted for in the spectra of the calibrator
stars. We also smoothed
the atmospheric models to the considerably lower resolution of the
calibrator spectra. The
comparison of the effective temperatures obtained with both methods
has a standard deviation of 109\,K, and we adopt that as the 1$\sigma$
uncertainty in the effective temperatures we derive in this
work.  Similarly, we adopt a five percent standard deviation in radius.
While the temperatures and radii we report in Table \ref{tab:temp}
overlap with only the cool end of the calibrator sample, inspection of
Figures \ref{fig:tempinterferometry} and
\ref{fig:radiusinterferometry} show no systematic trends. We performed
a Student's t test and found that the effective temperatures derived
with interferometry and with atmospheric fits are consistent with
belonging to the same sample to 0.89 significance.

We therefore conclude
that within the 1$\sigma$ uncertainties we adopt (109\,K for effective temperature
and five percent for radius) our method is capable of determining the
true effective temperature of a sample of stars in a statistical sense.
We propagate those uncertainties when using the values in Table  \ref{tab:temp}
to evaluate models of stellar structure and evolution and note that those
results (Section \ref{subsec:evolutionary}) should also be viewed as a
statistical treatment.

\begin{deluxetable}{lcccc}[h!]
\tabletypesize{\scriptsize}
\tablecaption{Comparison with Interferometric Calibration Sample \label{tab:interferometry}}
\tablehead{ \colhead{Star}              &
            \colhead{Temperature (K)}   &
            \colhead{   }               &          
            \colhead{Radius     }       \\
            \colhead{     }             &
            \colhead{Atm. fit}          &
            \colhead{Interferometry}    &
            \colhead{Atm. fit}          &
            \colhead{Interferometry}    }
\startdata
GJ 15 A  &  3631 & 3602$\pm$13  &   0.3849  &   0.3863$\pm$.0021 \\	   
GJ 105 A &  4823 & 4704$\pm$21  &   0.7688  &   0.7949$\pm$.0062 \\
GJ 205   &  3600 & 3850$\pm$22  &   0.6455  &   0.5735$\pm$.0044 \\        
GJ 338 A &  4147 & 3953$\pm$37  &   0.5326  &   0.5773$\pm$.0131 \\        
GJ 338 B &  4048 & 3926$\pm$37  &   0.5561  &   0.5673$\pm$.0137 \\
GJ 380   &  4019 & 4176$\pm$19  &   0.6923  &   0.6398$\pm$.0046 \\
GJ 412 A &  3644 & 3537$\pm$41  &   0.3888  &   0.3982$\pm$.0091 \\          
GJ 436   &  3448 & 3520$\pm$66  &   0.4476  &   0.4546$\pm$.0182 \\          
GJ 526   &  3644 & 3646$\pm$34  &   0.5164  &   0.4840$\pm$.0084 \\        
GJ 570 A &  4639 & 4588$\pm$58  &   0.7286  &   0.7390$\pm$.0190 \\          
GJ 581   &  3380 & 3487$\pm$62  &   0.3256  &   0.2990$\pm$.0100 \\        
GJ 687   &  3417 & 3457$\pm$35  &   0.4317  &   0.4183$\pm$.0070 \\        
GJ 699   &  3257 & 3238$\pm$11  &   0.1926  &   0.1869$\pm$.0012 \\          
GJ 702 B &  4305 & 4475$\pm$33  &   0.7301  &   0.6697$\pm$.0089 \\
GJ 725 A &  3453 & 3417$\pm$17  &   0.3538  &   0.3561$\pm$.0039 \\        
GJ 809   &  3662 & 3744$\pm$27  &   0.5536  &   0.5472$\pm$.0066 \\          
GJ 820 A &  4313 & 4399$\pm$16  &   0.6867  &   0.6611$\pm$.0048 \\          
GJ 820 B &  4052 & 4025$\pm$24  &   0.5862  &   0.6010$\pm$.0072 \\        
GJ 880   &  3613 & 3731$\pm$16  &   0.5770  &   0.5477$\pm$.0048 \\        
GJ 887   &  3691 & 3695$\pm$35  &   0.4751  &   0.4712$\pm$.0086 \\        
GJ 892   &  4734 & 4773$\pm$20  &   0.7984  &   0.7784$\pm$.0053 \\         
\enddata
\end{deluxetable}

\subsubsection{General Considerations Regarding the Atmospheric Fits} \label{subsubsec:atmconsiderations}
As a general trend we note that the BT-Settl spectra provide good fits
to the observed data. The quality of the fits is best noted in the
digital supplement to Figures \ref{fig:spectra1} and
\ref{fig:spectra2}, which show the data in its original unsmoothed
version. All fits pass the binary metallicity test, where the same
range of metallicity must be predicted for the two components of the
same binary system, to within about 0.1 dex.  The metallicity
  of two systems, GJ 22 ABC and GJ 1245 ABC is independently known
  from the isolated B component, and that information can be used as a
  test on our procedure. \citet{Rojas-Ayalaetal2012} report
  $[Fe/H]=-0.19$ for GJ 22 B based on infrared spectroscopy. Our
  procedure finds $[Fe/H]=-0.19$ for GJ 22 A, and $[Fe/H]=-0.10$ for
  GJ 22 C, thus validating results to no better than 0.1 dex. For GJ
  1245 AC \citet{Benedictetal2016} report $[Fe/H]=-0.04$. We obtain
  $[Fe/H]=-0.07$ for the A component and $[Fe/H]=-0.08$ for the C
  component. Despite this agreement, we notice that all 10 stars
  appear to have slightly sub-solar metallicity in our analysis. This
  feature could be real or it could be an artifact due to the boundary
  effect in the metallicity scaling of the model grid. For most
  temperature and gravity combinations the metallicity ([Fe/H]) ranges
  from -2.0 to 0.0, with only sporadic coverage up to +0.5. A random
  linear combination of model spectra (Section
  \ref{subsec:atmospheric}) is therefore likely to be biased towards
  lower metallicities even if in fact the metallicity is very close to
  solar. We therefore de-emphasize the absolute scaling of the
  metallicities in Table \ref{tab:temp} and focus on the broad
  agreement between the metallicities of components of the same
  system. The
models and the data show remarkable fine scale agreement in a line by
line basis down to the noise limit, particularly at wavelengths
  blue than 8,000\,\AA.

The fits to the gravity sensitive KI doublet at 7,700\,\AA $~$and NaI
doublet at 8,200\,\AA, and to the TiO bands starting at around
6,650\,\AA, 7,050\,\AA, and 7,590\,\AA are also generally
very good.  We note, however, that within the limitations of the
$Log\,g$ grid spacing there appears to be a slight bias
towards calculated surface gravities being higher than the fit
values, however without a finer grid it is impossible to determine
the significance of this tendency. The mean of the residuals
  of those fits in the sense of $Log\,g$ derived from the atmospheric
  fits minus that derived from radius and mass is -0.15. If we exclude the
  poorly modeled stars GJ 1245 A and C the mean of the residuals becomes
  -0.09. There is a possibility that this discrepancy is due to the radii derived
  from atmospheric models being under-estimated. Given the
  check on the radius methodology from the interferometric data,
  we believe that such an effect, if present, is small and well within the
  uncertainties of the derived radii because we see no systematic
  trends on Figure \ref{fig:radiusinterferometry}. At the temperatures we
consider here surface gravity has little effect on overall
spectrum morphology except for altering specific gravity sensitive
features such as the KI and NaI lines. When calculating the
overall best fit model (Section \ref{subsec:atmospheric}) the
wavelength coverage of these lines may be too small to
meaningfully influence the fit.  A direct comparison of the
morphology of the KI and NaI doublets between model and spectra
may be a better indication of surface gravity, however such
comparison would require assumptions in temperature and
metallicity.  We weighed both approaches and decided to keep the
single fit approach because it is a good compromise
between diagnosing both effective temperature and surface gravity to
reasonable accuracy. We note that the only star with a large
discrepancy between the atmospheric model predicted surface
gravity and the surface gravity calculated from radius and mass is
GJ 1245 C, and in that case the fits to the gravity indicators,
particularly the KI doublet, is also poor.

The quality of the fits deteriorates at lower temperatures, as is
evident in the fits for GJ 469 B (3,134\,K), GJ 1245 A
(2,960\,K), and particularly GJ 1245 C (2,611\,K). The
problem could be in part due to the intrinsic difficulty of modeling
spectra at cooler temperatures where molecular species and grains
become more important, but also due to greater sensitivity to
temperature itself. The overall slope of M dwarf spectra increases
rapidly as a function of temperature at temperatures $\lesssim$
3,100\,K, and finding a simultaneous fit to the blue and red parts of
the spectrum therefore requires a finer grid.

We also note that the depths of individual lines in the red part of
the model spectra beyond 8,000\,\AA$~$ seem to be too shallow while
the general shape of the spectrum is still a good match. In other
words, the model spectra are smoother than the observed spectra. The
fact that we still see a line to line match at the smallest scale, as
best seen in the high resolution online supplement to Figures
\ref{fig:spectra1} and \ref{fig:spectra2}, indicates this
discrepancy is not due to noise in the observed spectra. The effect is
also distinct from fringing, which is a larger scale feature that
tends to alter the spectra in a step-like manner; even within fringed
regions, the shallower individual line depths are still present.

Our ability to derive accurate effective temperatures and radii
is in large part due to the choice of wavelength region to study, and
does not necessarily speak to the adequacy of the BT-Settl models
in other wavelength ranges. Even in our case the first 118\,\AA $~$and the last
277\,\AA $~$of the observed spectra were omitted from the fit due to considerable
deviations between observations and models, as shown in the shaded regions
of Figures \ref{fig:spectra1} and \ref{fig:spectra2}. The blue region in
particular shows residuals of up to 30 percent. It is also well known
that incomplete oppacities create problems in the near infrared, particularly
in the J band, and that those discrepancies hinder the determination of
bolometric flux from model spectra alone. \citet{Baraffeetal2015} also note that
TiO line lists are still incomplete and the incompleteness can cause
problems particularly at higher resolutions, however we do not notice
higher than usual residuals to TiO bands in Figures \ref{fig:spectra1} and
\ref{fig:spectra2}. \citet{Rajpurohitetal2018} also note problems with line
widths at higher resolutions, which again do not seem to be problematic
at the resolution of this work, as best visualized in the high resolution supplement
to Figures  \ref{fig:spectra1} and \ref{fig:spectra2}. Overall, the red optical and
very near infrared region we study here, from 6,600\,\AA $~$to 9,850\,\AA,
is well modeled in this temperature regime, and the comparison with
interferometric results (Section \ref{subsubsec:interferometry}) validates
the parameters we derive from atmospheric models, therefore allowing
us to use them as the comparison standard for testing models of stellar
structure and evolution.
 
\section{Testing Evolutionary Models} \label{subsec:evolutionary}
\begin{figure}[h!]
\gridline{\fig{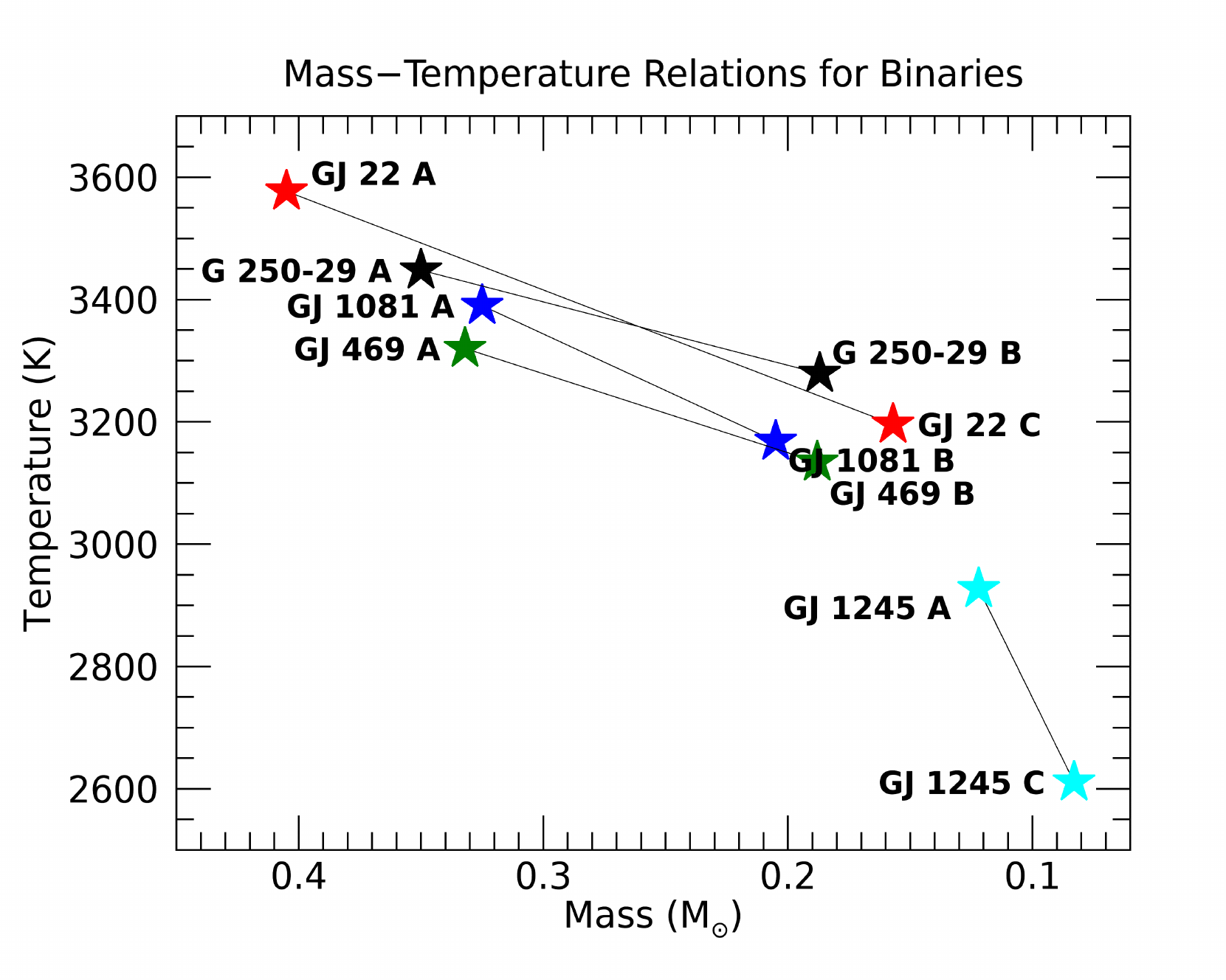}{0.4\textwidth}{(a)}}
\gridline{\fig{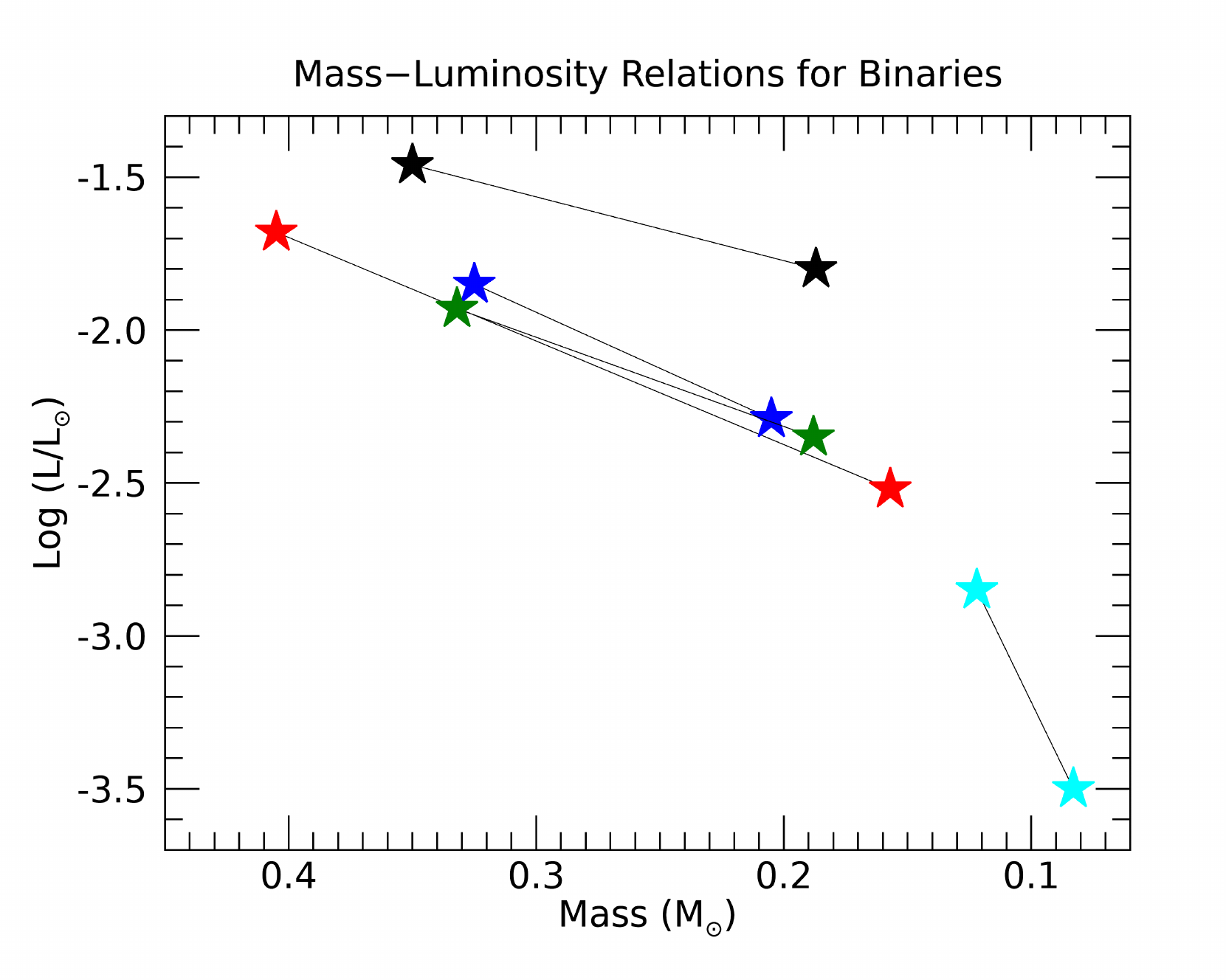}{0.4\textwidth}{(b)}}
\gridline{\fig{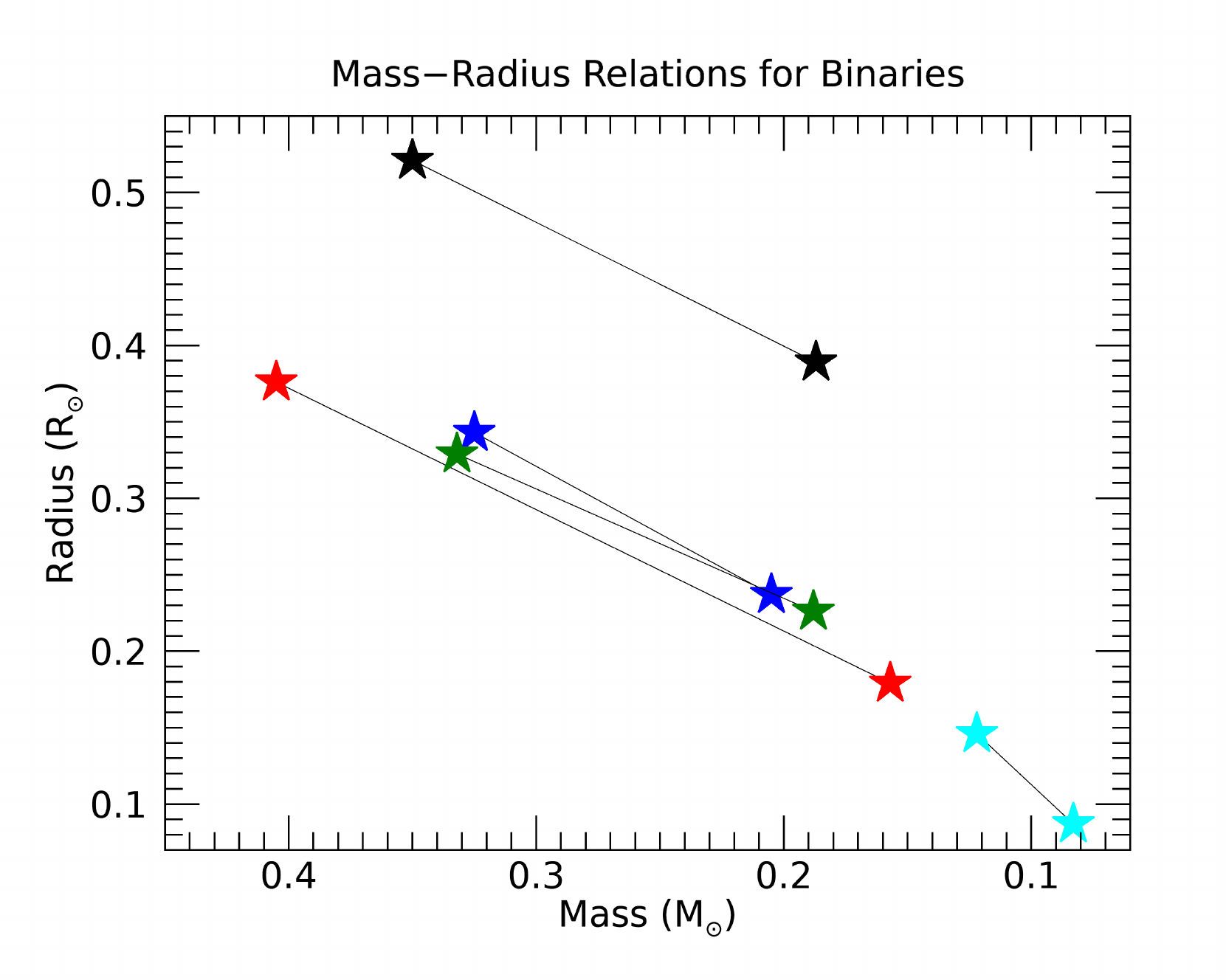}{0.4\textwidth}{(c)}}
\caption{\scriptsize Temperature, luminosity, and radius as a
    function of mass for the stars in the sample.  Components of the
    same binary are connected with thin lines and color coded. Panels
    b and c follow the same color coding as a.  Valid evolutionary
    models must produce isochrones that can replicate these quantities
    for both components of a given binary
    simultaneously.  \label{fig:masstemp}}
 \end{figure}

Figure \ref{fig:masstemp} shows the distributions of temperatures,
luminosities, and radii as functions of mass for the observed
sample. An ideal set of evolutionary predictions would be able to
replicate these values while respecting the coevality of stars in the
same system. Here we focus on five evolutionary model suites that are
commonly used to study low mass stars: the Dartmouth models
\citep{Dotteretal2008}, the MIST models
\citep{Choietal2016,Dotteretal2016}, the models of
\citet{Baraffeetal2015}, the PARSEC models \citep{Bressanetal2012},
and the YaPSI models \citep{Spadaetal2017}.

Figures \ref{fig:evol1}, \ref{fig:evol2}, and \ref{fig:evol3} show
evolutionary tracks interpolated to the masses of each star in
temperature, luminosity, and radius. Each panel shows
the model predictions for the two components of a star system with the
results from Table \ref{tab:temp} overlaid as shaded regions encompassing
the observational
constraints from this study. Only the \citet{Baraffeetal2015} models
and the PARSEC models incorporate cool enough atmospheres to model the
properties of the coolest star in the study, GJ 1215 C.

Table \ref{tab:matches} summarizes the graphical results of Figures
\ref{fig:evol1}, \ref{fig:evol2}, and \ref{fig:evol3} by tabulating
the instances in which a given model can accurately predict the
observed properties assuming either main sequence ages or
pre-main sequence ages. For the purpose of this study we define
  the zero age main sequence as the point of maximum radius
  contraction. In Table \ref{tab:matches} a fully self-consistent
  model match in the sense that a model can predict all three
  fundamental parameters for both stars in a system in a coeval manner
  is marked with the symbol \checkmark \checkmark \checkmark. That
  only happens in the case of the GJ 22 AC system (Section
  \ref{subsec:gj22ac}). The symbols ``YY'' and ``Y'' denote when a
  parameter is correctly modeled if the system is young (pre-main
  sequence) while respecting or not respecting coevality,
  respectively. The interpretation of these pre-main sequence matches
  must be done with caution. From the shape of the evolutionary tracks
  for luminosity and radius in Figures \ref{fig:evol1},
  \ref{fig:evol2}, and \ref{fig:evol3} it is nearly always possible to
  find a pre-main sequence solution that falls in the desired
  parameter space for a short time in the system's evolution. While we
  cannot discard these instances as valid matches for pre-main
  sequence systems, it is unlikely that many of the five star systems
  we study lie in such a specific narrow range of the pre-main
  sequence. We note also that none of the systems have a fully
  self-consistent pre-main sequence solution.

We now discuss topics related to individual
model sets and save a general  discussion on
how well these models work as a whole to Section \ref{sec:conclusions}.

\subsection{Dartmouth Stellar Evolution Database} \label{subsubsec:dartmouth}
The Dartmouth Stellar Evolution Database and its stellar evolution
code\footnote{http://stellar.dartmouth.edu/models/}
\citep{Dotteretal2008} is an older code that has been periodically
updated to provide results in several photometric systems and increase
functionality. Its web interface allows for the easy production of
isochrones and evolutionary tracks interpolated to any mass, age, and
metallicity within their parameter ranges. We used this web interface
to produce the results shown in Figures \ref{fig:evol1},
\ref{fig:evol2}, and \ref{fig:evol3}. One particular limitation is
that it does not include ages younger than 1\,Gyr, and so only main
sequence stars can be modeled.  This age limitation also excludes the
zero age main sequence, as shown most clearly in the radius plots
where the other evolutionary models show radius minima.

The Dartmouth models use atmospheric boundary conditions based on the
NextGen atmospheric models generated with the PHOENIX radiative
transfer code \citep{Hauschildtetal1999a, Hauschildtetal1999b}. These
model atmospheres use the older solar abundances of
\citet{GrevesseAndSauval1998}, which have since been revised several
times, as discussed in \citet{Allardetal2013}.  Interestingly, this
choice of older atmospheric models and older solar metallicities does
not seem to drastically affect results when compared to other models.
This is in contrast to the effect of using the older solar abundances
of \citet{GrevesseAndSauval1998} in atmospheric models, which can
cause a noticeable difference in predicted effective temperatures
\citep{MannGaidosAndAnsdell2013}.  We note the wide discrepancy in
temperature in the case of GJ 1245 A, where our results show an upper
bound on the temperature of 3036\,K and the model predicts 3,200\,K.
A detailed treatment of metallicity becomes more important at lower
temperatures as molecular species begin to form and greatly increase
the opacity. Therefore, the choice of solar metallicities could be the
cause of the temperature discrepancy for GJ 1245 A, which is
significantly cooler than the other stars, except for GJ 1245 C.

\subsection{MESA Isochrones and Stellar Tracks $-$ MIST} \label{subsubsec:mist} 
The MIST models\footnote{http://waps.cfa.harvard.edu/MIST/}
\citep{Choietal2016,Dotteretal2016} are an application of the Modular
Experiments in Stellar Astronomy
(MESA)\footnote{http://mesa.sourceforge.net/index.html} code that
tabulates evolutionary tracks for a wide range of stars using
solar-scaled metallicities. The solar metallicity zero points are set to the
values adopted in \citet{Asplundetal2009}. MESA is a large, open
source comprehensive project that preserves a wide range of freedom in
the input parameters for the stellar evolution code, so choosing
the adequate parameters for a particular model grid is in itself a
complex scientific task.

The evolutionary tracks shown in Figures \ref{fig:evol1},
\ref{fig:evol2}, and \ref{fig:evol3} were produced using the MIST web
interpolator. A feature that readily stand out are what appear to be
pulsations, with a period in the order
of hundreds of millions of years. No other model shows that behavior.
The pulsations are the strongest in the 0.32\,M$_{\odot}$ to
0.35\,M$_{\odot}$ mass range, which is close to the mass where stars
become fully convective. We discuss issues relating to the onset of full convection
in detail in Section \ref{subsec:kissing}.

One of the distinct advantages of the MESA/MIST approach is the
ability to generate new model grids based on different input
parameters with relative ease and with minimal knowledge of the inner
workings of the code. In that sense it may be possible to produce a
different MESA implementation that is calibrated to a narrower set of stars
and provides a better match to those observations.

\subsection{Models of Baraffe et al., (2015)}\label{subsubsec:baraffe}
The evolutionary models of \citet{Baraffeetal2015} are the latest in a
long tradition of evolutionary models that are not only stellar, but
also bridge the stellar-substellar boundary and model the brown dwarf
domain.  This family of models has incorporated various versions of
the PHOENIX model atmospheres as a boundary condition, and this latest
installment incorporates the BT-Settl atmospheres used in this work
(Sections \ref{subsec:atmospheric} and
\ref{subsubsec:atmconsiderations})\footnote{ The evolutionary models
  of \citet{Baraffeetal2015} are sometimes erroneously referred to as
  ``the BT-Settl models''. While it is true that they incorporate the
  BT-Settl atmospheric models as a boundary condition and that the
  authors of both atmospheric and internal models work in close
  collaboration in this case, clarity demands that the term
  ``BT-Settl'' be reserved for the atmospheric models.}.

The BT-Settl atmospheric models are arguably the most advanced model
atmospheres used as a boundary condition for any of the evolutionary
models we studied.  Further, the fact that the \citet{Baraffeetal2015}
models incorporate the same atmospheres we used to derive fundamental
parameters leads us to expect a somewhat better agreement between
those parameters and the model predictions. Yet, their results are
mixed and not qualitatively better than the models that use other
atmospheric boundary conditions (Dartmouth and MIST). The same can be
said for the YaPSI models, which also incorporate the BT-Settl
atmospheres.  This consideration again suggests that any mismatches
may be indicative of deeper theoretical discrepancies independent of
the choice of atmospheric boundary conditions.

One advantage of the \citet{Baraffeetal2015} models is that they
include significantly lower temperatures, and along with PARSEC are
the only models discussed here that can model GJ 1245 C at
2,611$\pm$109\,K.  However as seen in Figure \ref{fig:evol3}, the
temperature predictions would only agree to the inferred value if the
GJ 1245 system is young, with an age ranging from 250\,Myr to
800\,Myr, and that would then be in disagreement with the predictions
for luminosity. While both components of GJ 1245 exhibit H$\alpha$
emission (Figure \ref{fig:spectra2}, best seen in the high resolution
digital supplement) such emission is common in this very low mass
regime (note also H$\alpha$ emission in GJ 22 C and GJ 1081 B), and
does not necessarily indicate youth \citep[e.g.,][]{Browningetal2010}.

\subsection{The PARSEC Models}\label{subsubsec:parsec}
\begin{figure*}[h!]
  \gridline{\fig{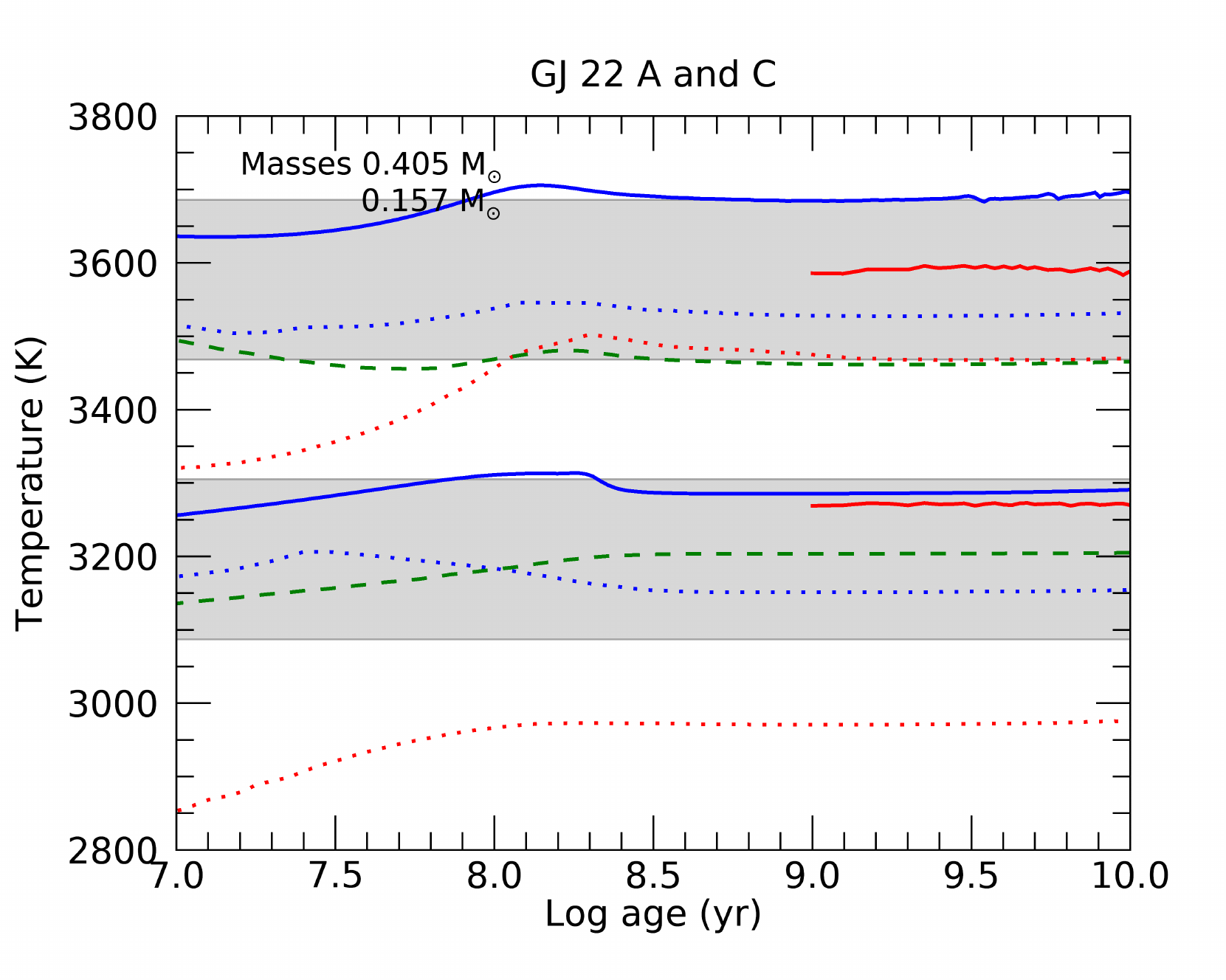}{0.39\textwidth}{}
            \fig{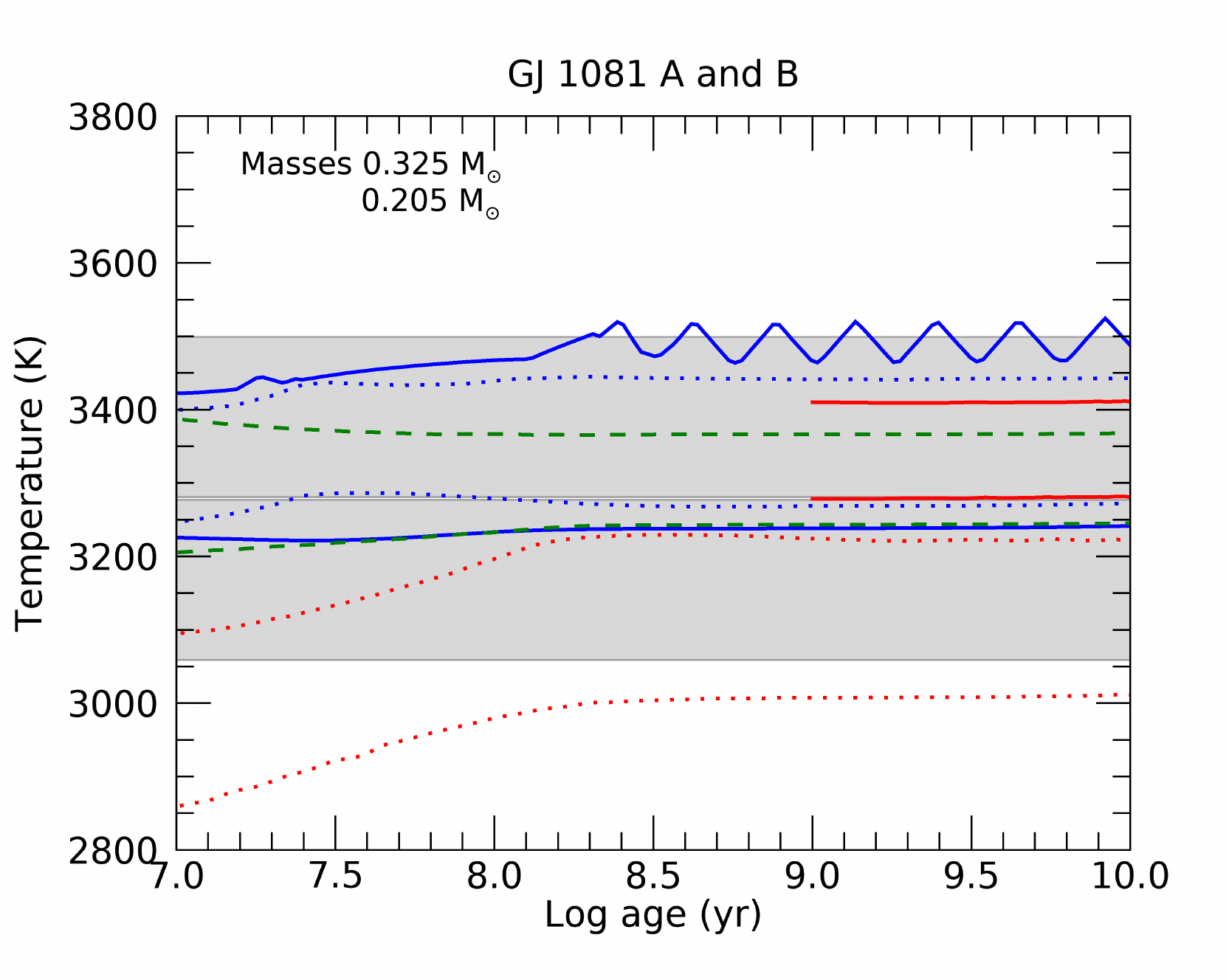}{0.39\textwidth}{}
           }
  \gridline{\fig{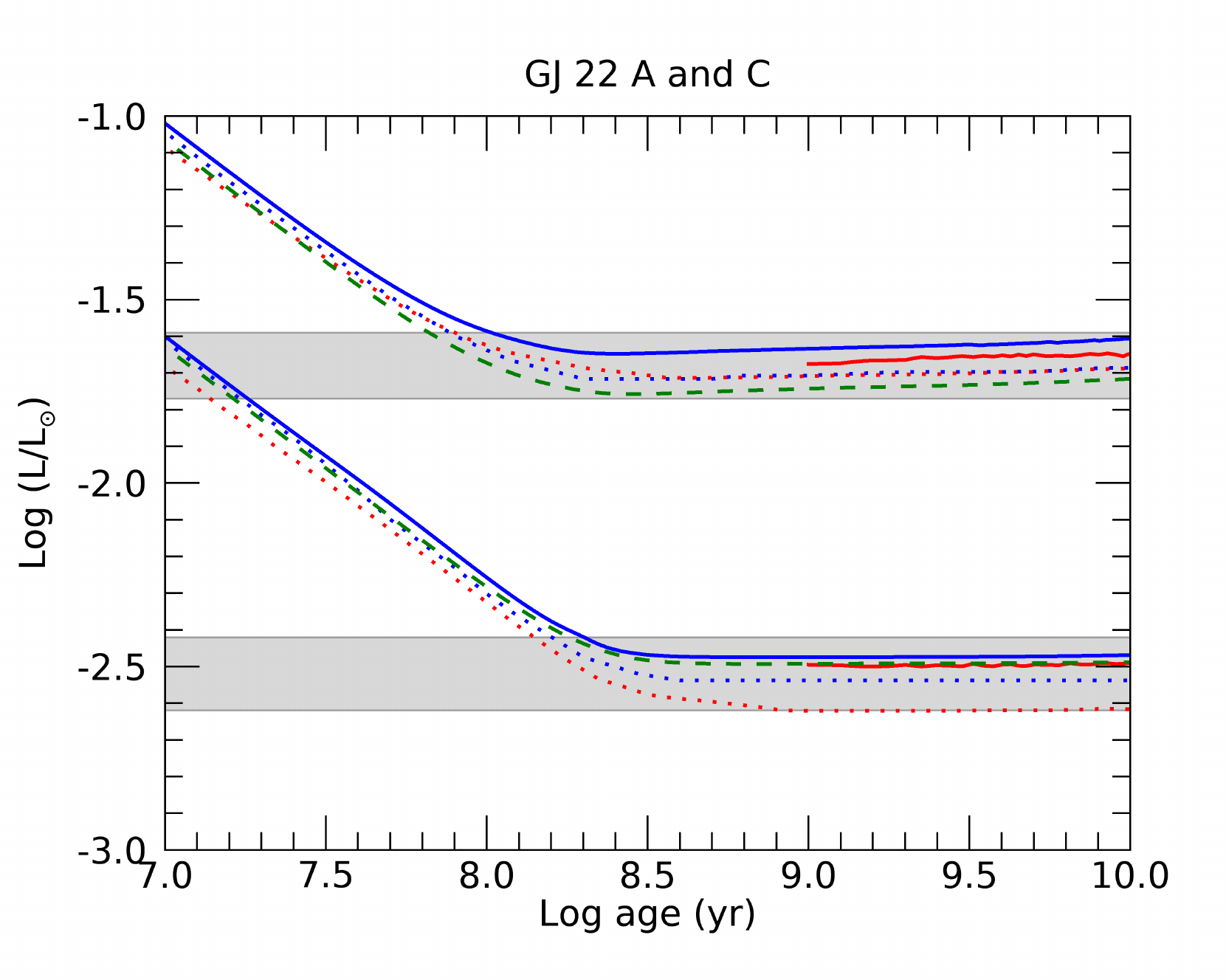}{0.39\textwidth}{}
            \fig{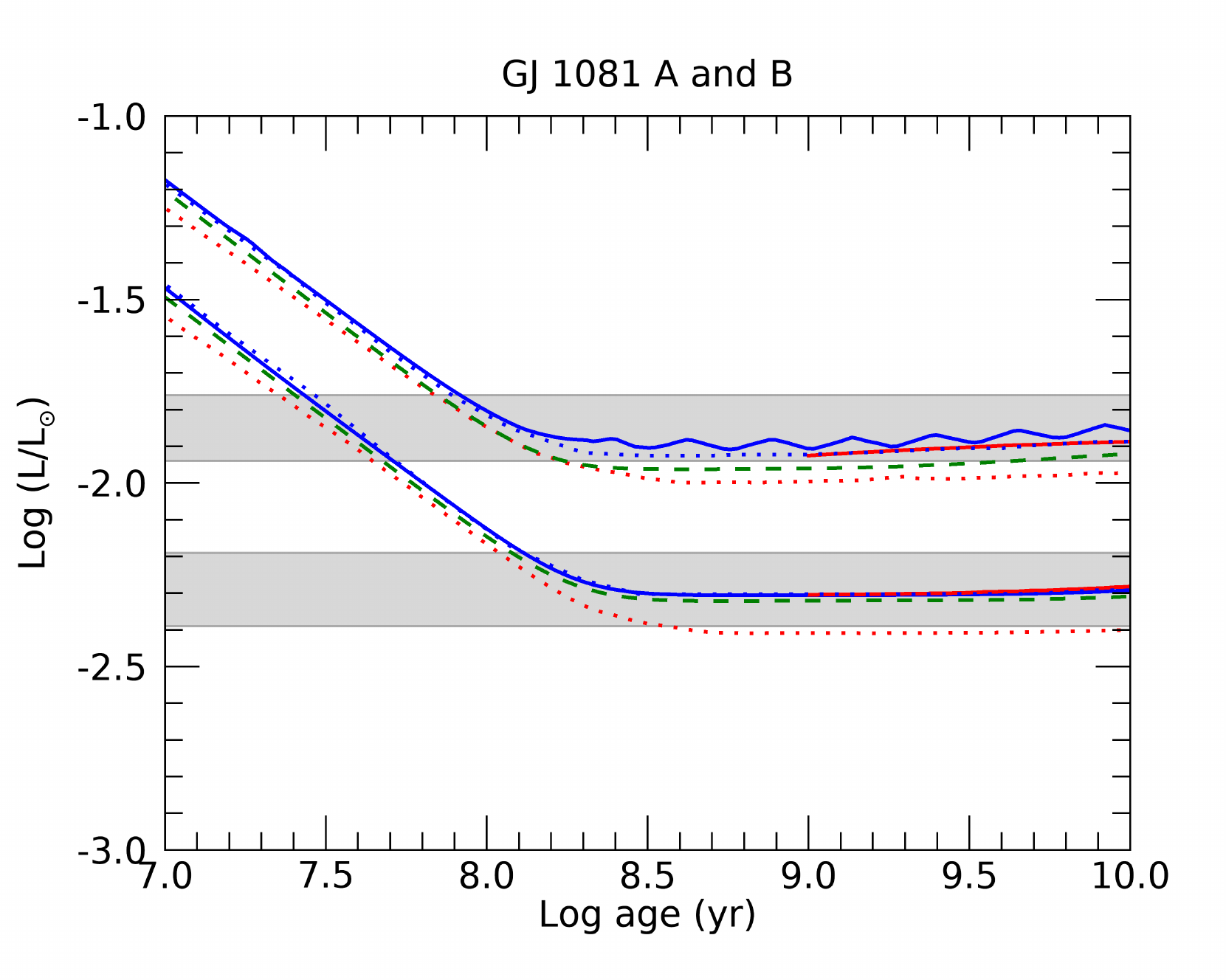}{0.39\textwidth}{}
           }
  \gridline{\fig{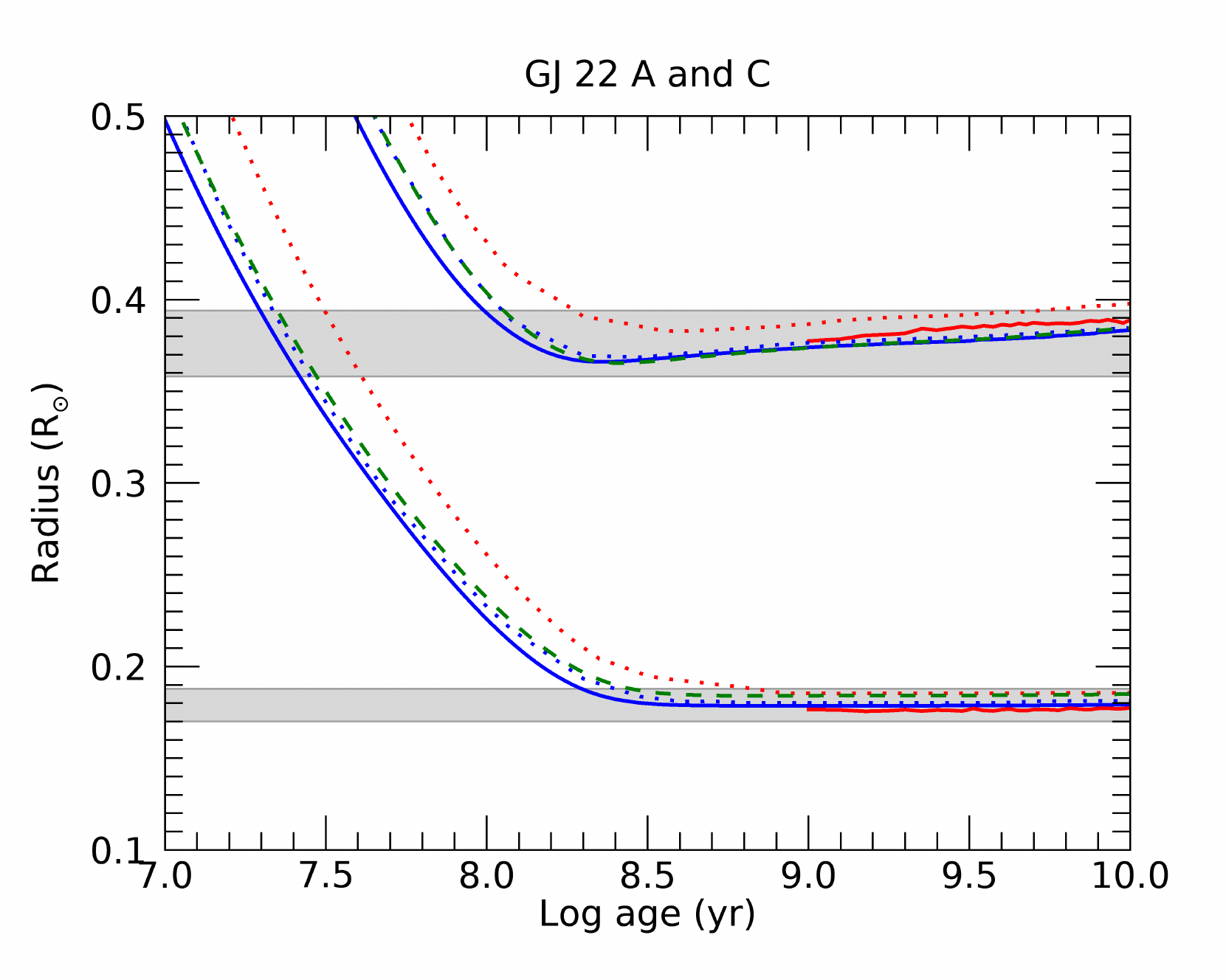}{0.39\textwidth}{}
            \fig{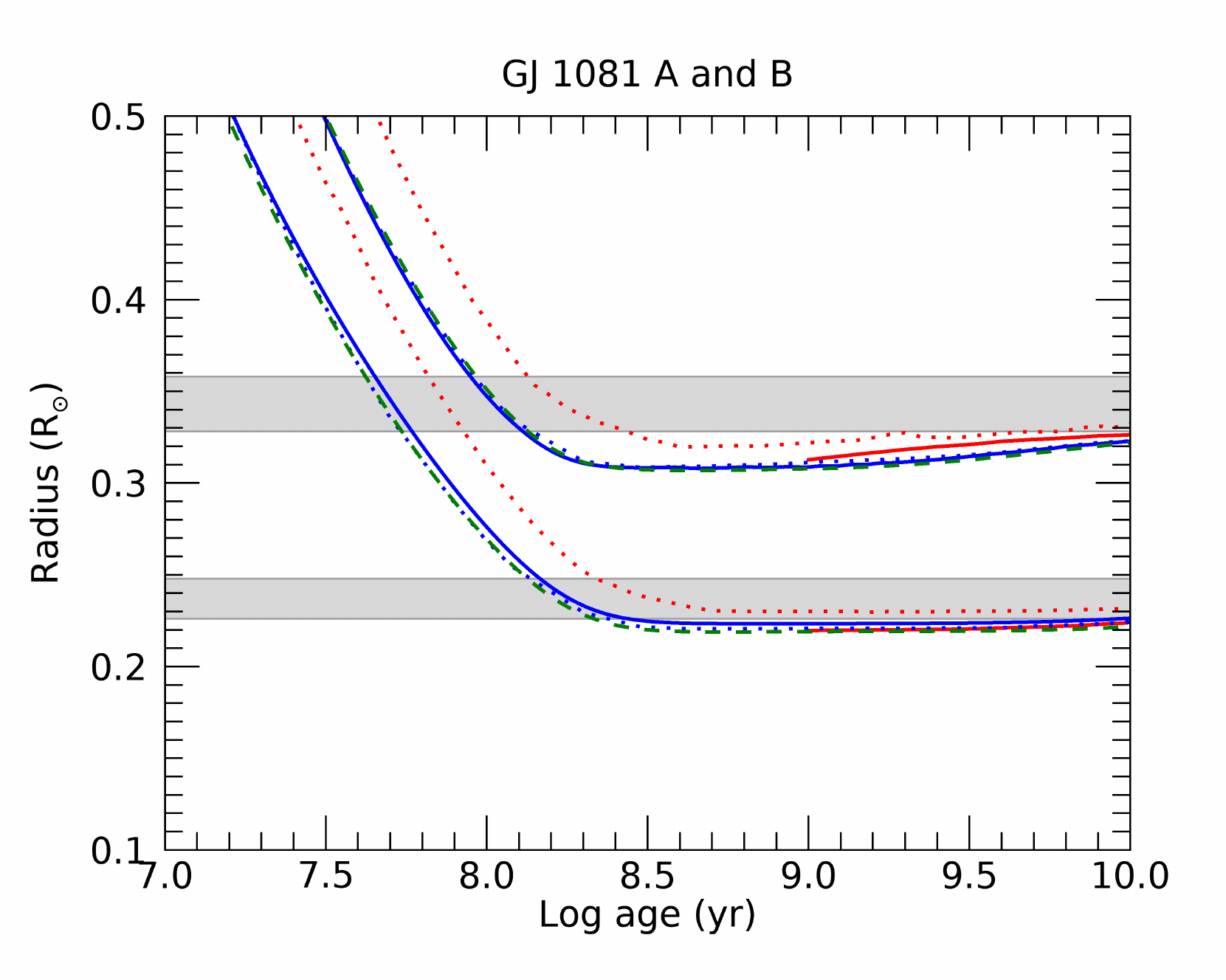}{0.39\textwidth}{}
           }
  \caption{\scriptsize  Evolutionary
    plots for the GJ 22 (left column)
    and GJ 1081 (right column) systems. The line style and color
    scheme is as follows: solid red line for Dartmouth plots, solid
    blue line for MIST plots, blue dotted lines for the Baraffe
    models, red dotted lines for the PARSEC models, and green dashed
    lines for the YaPSI models. Two tracks are shown for each model,
    corresponding to the primary and secondary components. The shaded
    areas show the uncertainties inferred from the atmospheric fits
    (Table \ref{tab:temp})See Section \ref{subsec:gj22ac} for a discussion
    of the metallicity of GJ 22 AC. \label{fig:evol1}}
\end{figure*}

\begin{figure*}[h!]
  \gridline{\fig{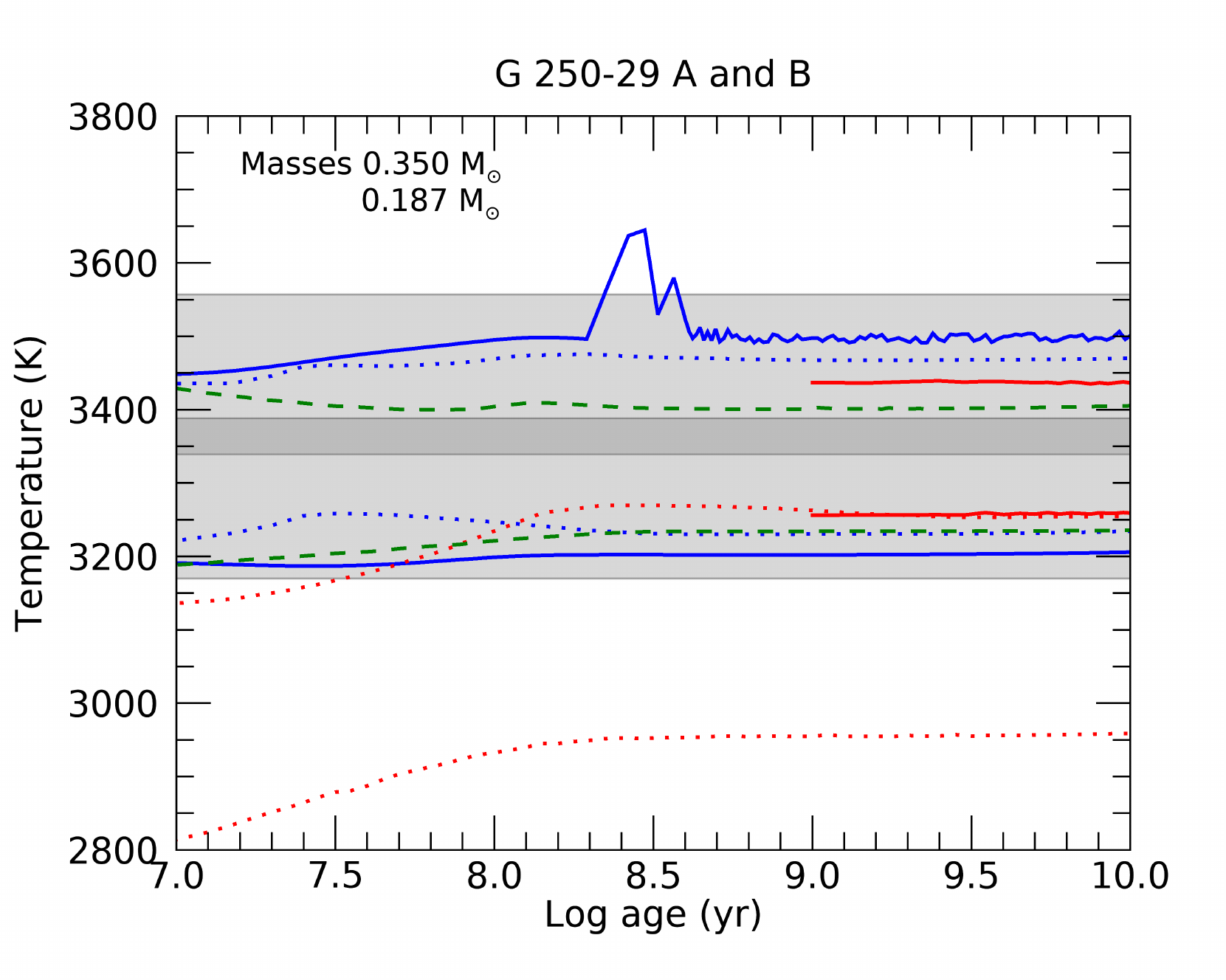}{0.45\textwidth}{}
            \fig{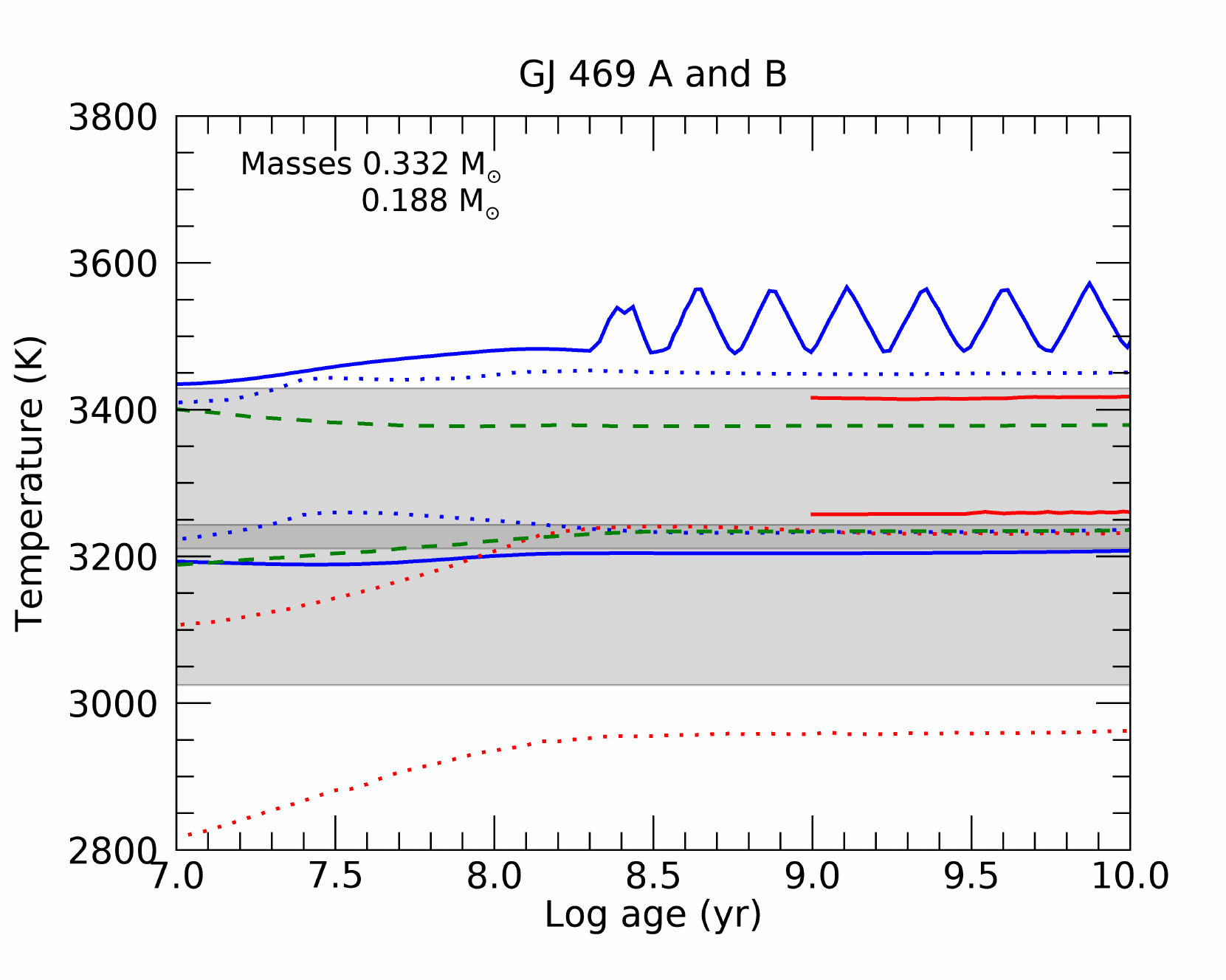}{0.45\textwidth}{}
           }
  \gridline{\fig{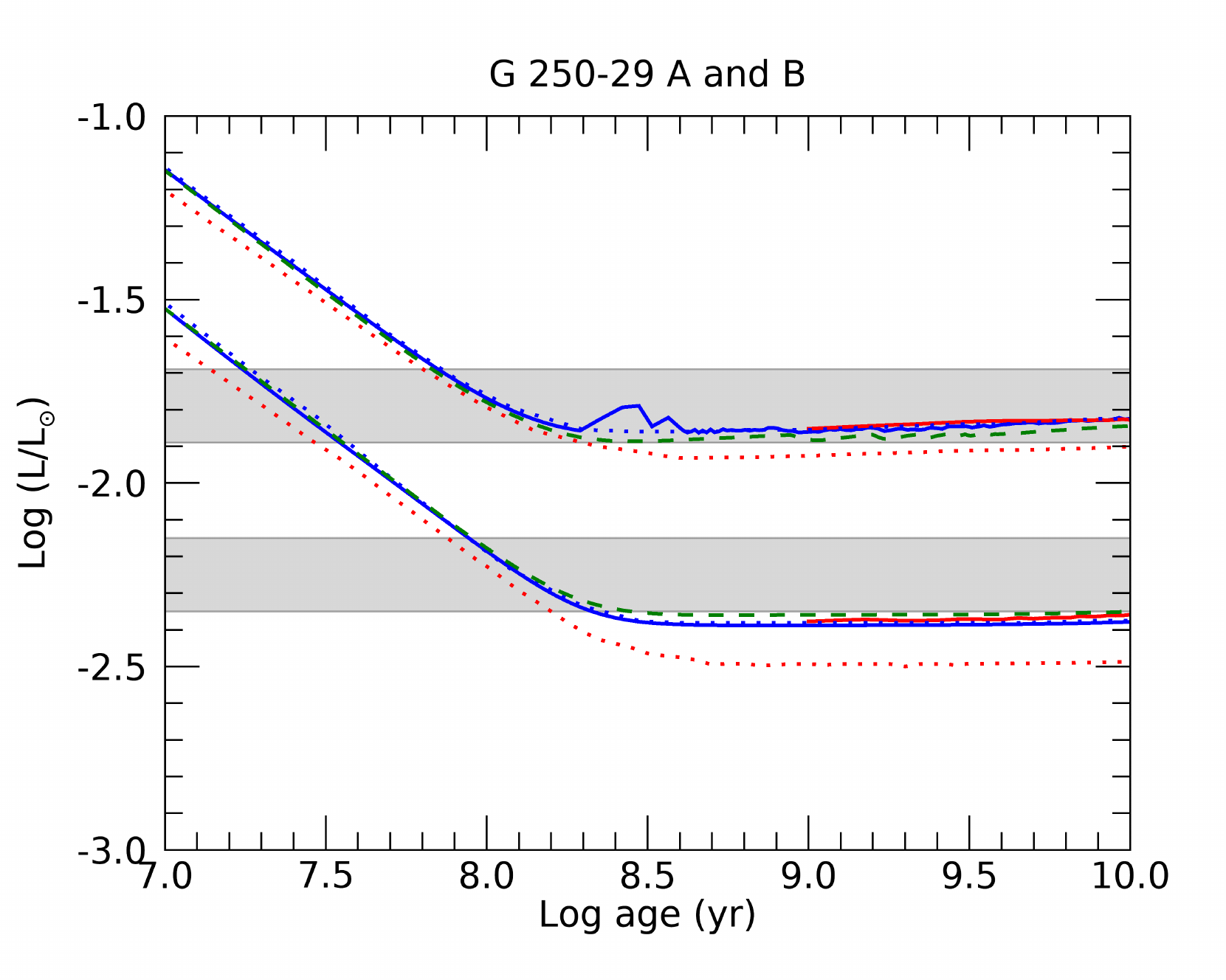}{0.45\textwidth}{}
            \fig{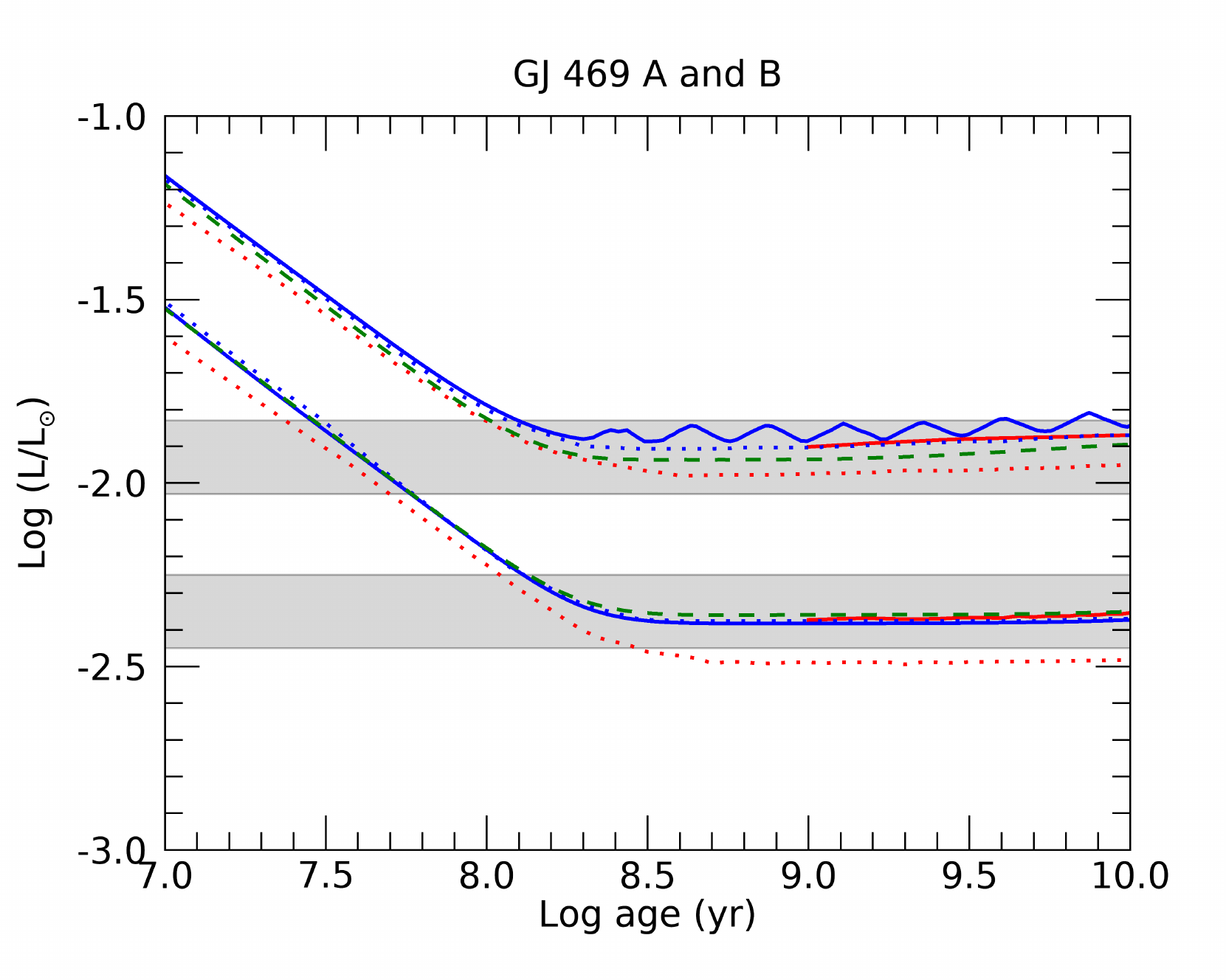}{0.45\textwidth}{}
           }
  \gridline{\fig{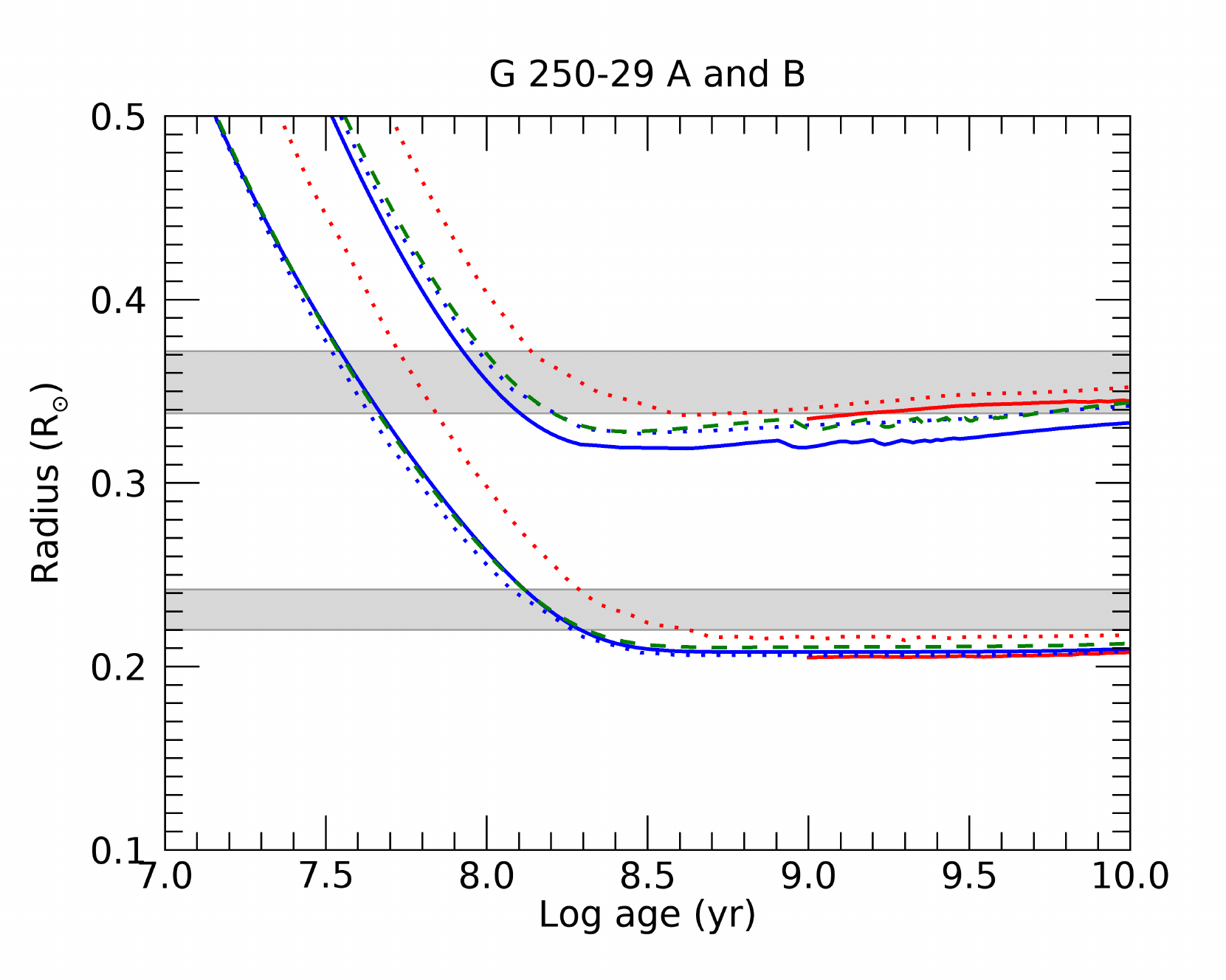}{0.45\textwidth}{}
            \fig{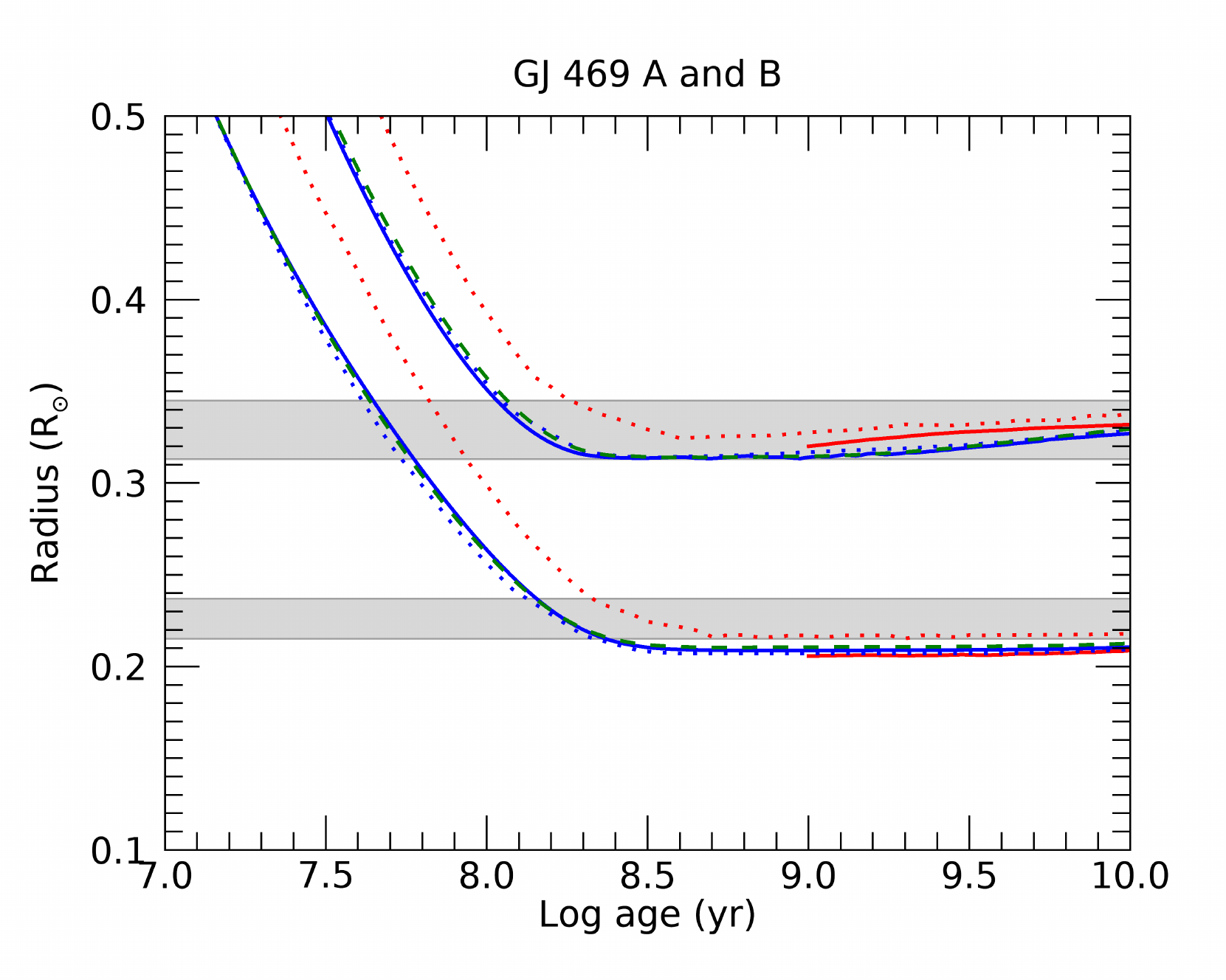}{0.45\textwidth}{}
           }
  \caption{\scriptsize Same as Figure \ref{fig:evol1} for the G 250-29 and GJ 469 systems. \label{fig:evol2}}
\end{figure*}

\begin{figure}[h!]
  \gridline{\fig{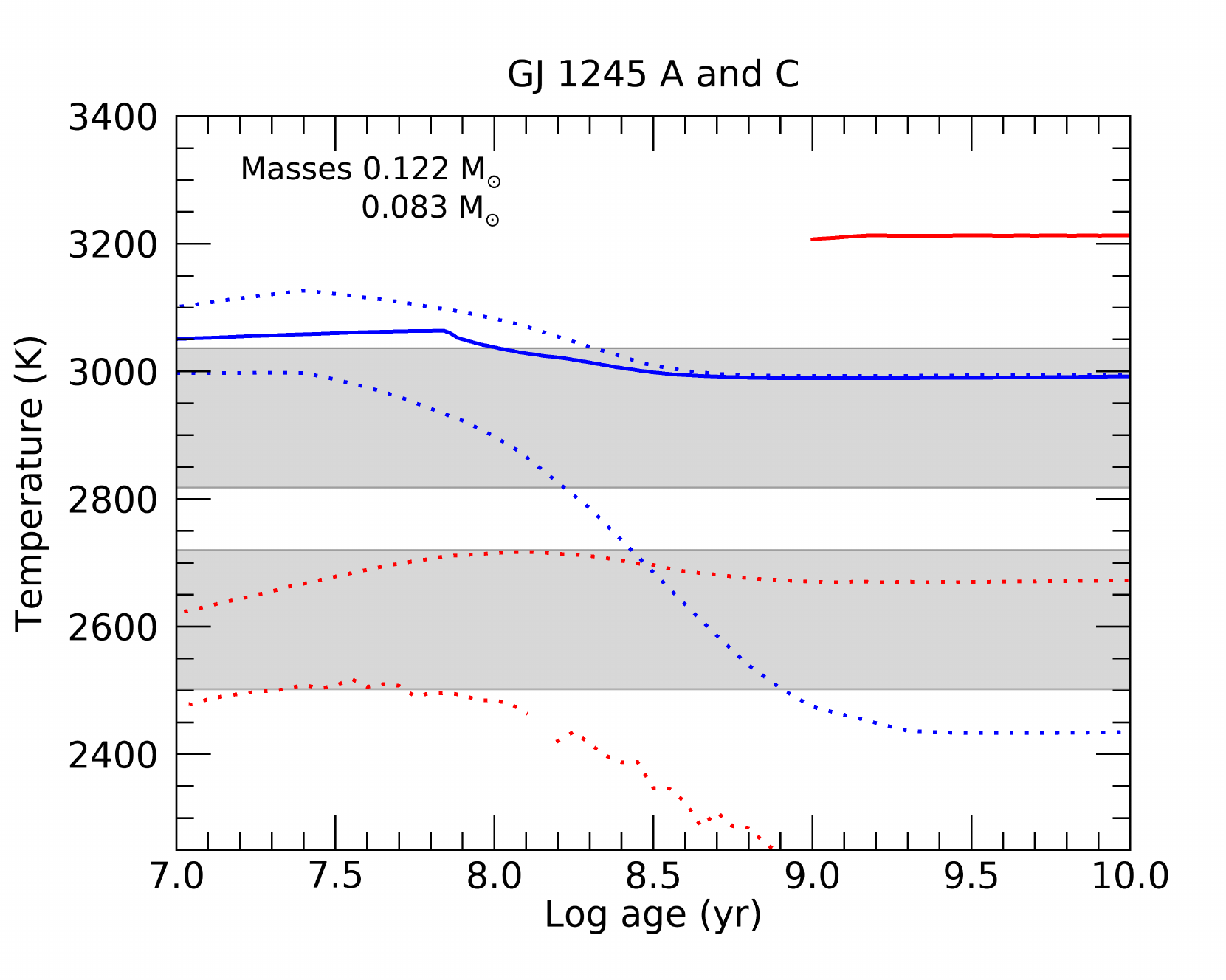}{0.45\textwidth}{}
           }
  \gridline{\fig{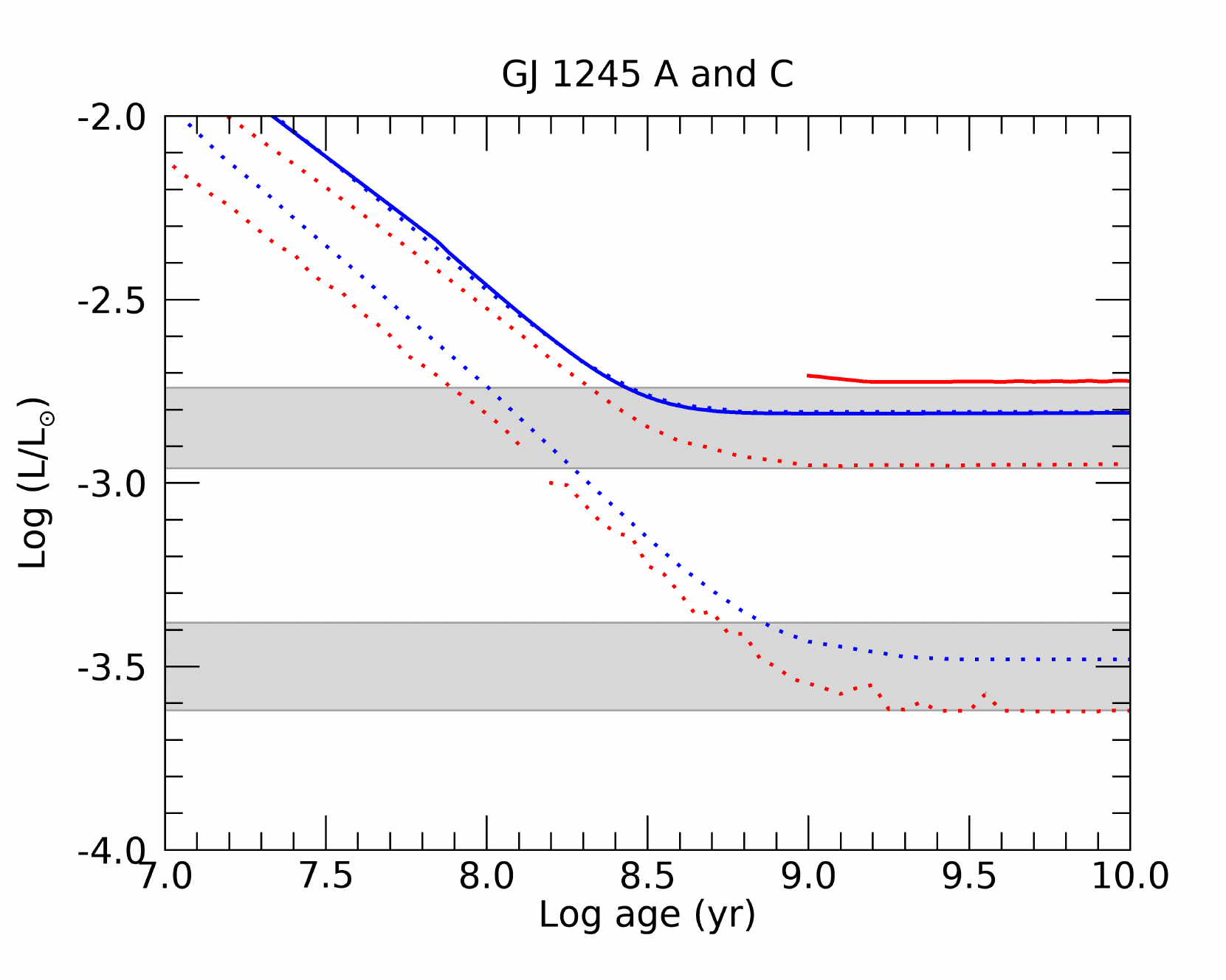}{0.45\textwidth}{}
           }
  \gridline{\fig{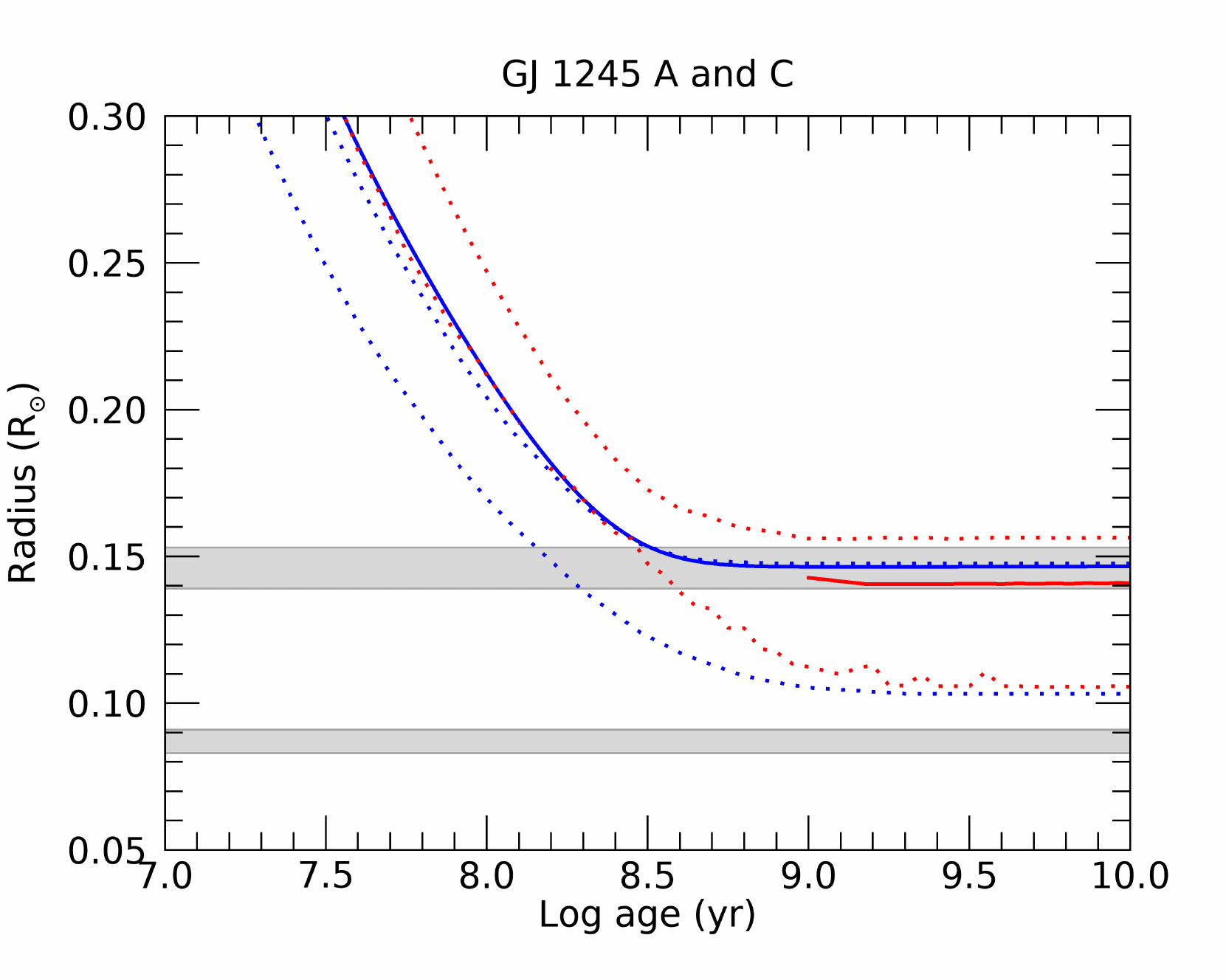}{0.45\textwidth}{}
           }
  \caption{\scriptsize  Same as Figure \ref{fig:evol1} for the GJ 1245 system.
    Only the Baraffe (blue dotted lines) and the PARSEC (red dotted lines) models reach cool enough temperatures
    to model GJ 1245 C.\label{fig:evol3}}
\end{figure}

The PAdova-TRieste Stellar Evolution Code \citep{Bressanetal2012} is a
versatile family of codes that over the years has developed specific
treatments for different regions of the HR
diagram. \citet{Chenetal2014} updated the code for the specific
treatment of the lower main sequence. The code allows the choice of a
wide range of parameters including a well populated metallicity grid
and has a convenient web
interface\footnote{http://stev.oapd.inaf.it/cgi-bin/cmd}.  The code
uses the BT-Settl atmospheres (Section \ref{subsec:atmospheric}) as
the boundary condition, albeit with the older metallicities of
\citet{Asplundetal2009}. \citet{Chenetal2014} calibrates the model
using the main sequences of several young and intermediate age star
clusters, and obtains remarkably good results from a populations
perspective in the sense that their isochrones are a good
representation of the cluster's main sequence. However in such a
comparison masses are treated indirectly in the sense that unless the
cluster's initial mass function is precisely known mass becomes a free
parameter for the color-magnitude fit. When masses become a fixed
parameter we find that the PARSEC models 
are systematically too cold. Their temperatures tend to be about 200\,K to 300\,K lower
than our inferred temperatures, as shown in the top panels of Figures
\ref{fig:evol1}, \ref{fig:evol2}, and \ref{fig:evol3}. While
  their radius predictions are in range with the other models, the
  predicted luminosities are accordingly lower. We note however that
if these mismatches could be fixed while preserving PARSEC's ability to
model populations in the HR diagram it would become a powerful tool.

\subsection{The YaPSI Models}\label{subsubsec:yapsi}
The Yale-Postdam Stellar Isochrones\footnote{http://www.astro.yale.edu/yapsi/}
\citep[YaPSI,][]{Spadaetal2017}
are an adaptation of the Yonsei-Yale family of codes
modified to emphasize the physics of low mass stars. Their approach
also relies on BT-Settl boundary conditions. Tow distinguishing
factors are an extremely fine mass grid and the availability of a
Markov Chain Monte Carlo grid interpolator available for download.
The ability to do
fine grid interpolation is useful when testing boundary cases, such
as the transition to full convection (Section \ref{subsec:kissing}). The authors tested the
YaPSI models using the mass-luminosity relation of \citet{Benedictetal2016}
(Section \ref{sec:obs}),
the predecessor to our study that provided the dynamical masses we use here.
Benedict et al. provide only $V$ and $K$ magnitudes for individual components,
as opposed to the fundamental parameters we provide here. \citet{Spadaetal2017}
show that the YaPSI models as an ensemble do a good job of replicating the color-magnitude
diagram of \citet{Benedictetal2016}, albeit with wide dispersion
about the model sequence. Our tests show that the YaPSI models do
comparatively well. Table \ref{tab:matches}
  shows that the YaPSI models do a slightly better job than the other
  models in predicting the parameters of individual stars, however they
  still lack the self-consistency necessary to fully match systems other than
  GJ 22. One drawback is that their lower
mass limit is 0.15\,M$_{\odot}$, and so they cannot model GJ 1245 A and C.

\section{Discussion} \label{sec:discussion}

With the exception of the PARSEC models most of the evolutionary
tracks shown in Figures \ref{fig:evol1}, \ref{fig:evol2}, and
\ref{fig:evol3} could be reconciled with the BT-Settl predictions
(Table \ref{tab:temp}) if the uncertainties in temperature were
increased by an additional 50\,K on each side and if uncertainties in
radius were about doubled. In that sense it may well be that our
expectations of evolutionary models are simply higher than what is
possible, and we should conform to temperature predictions no more
accurate than 300\,K. That realization would be unfortunate because M
dwarf effective temperatures and radii are routinely expressed to much
smaller uncertainties \citep[e.g.,][]{Dieterichetal2014, Mannetal2015}
and as discussed in Section \ref{subsubsec:interferometry} can indeed
be determined to better precision.

\subsection{Radius Inflation}\label{subsec:radius}
Several observational studies that measure M dwarf radii using
eclipsing binaries or optical interferometry indicate that the radii
of M dwarfs tend to be larger than those predicted by stellar
structure models \citep[e.g.,][]{Torresetal2010, Boyajianetal2012,
  FeidenAndChaboyer2012}. This trend is the so called radius inflation
problem. Most hypothesized explanations for the discrepancy involve
the interaction of magnetic fields with stellar matter.  We find that
radius inflation is indeed a significant problem, with more than half
of model predictions resulting in radii that are two small, as shown
in Table \ref{tab:radius}. We find only three instances of a radius
being {\it over} predicted, and those involve the GJ 1245 AC system,
which has proven more difficult to model due to its very low mass.

Inspection of Figures \ref{fig:evol1}, \ref{fig:evol2}, and
\ref{fig:evol3} and Table \ref{tab:matches} shows that radius
inflation is the leading reason why model predictions do not achieve
the desired self-consistent solutions for binary systems other than GJ
22. None of the models we test here include the effects of magnetism.
If radius inflation is indeed due to magnetic effects that would
provide a natural explanation of the problem.

\subsection{GJ 22 AC, A Well Behaved Metal Poor System?}\label{subsec:gj22ac}
\begin{figure}[h!]
  \gridline{\fig{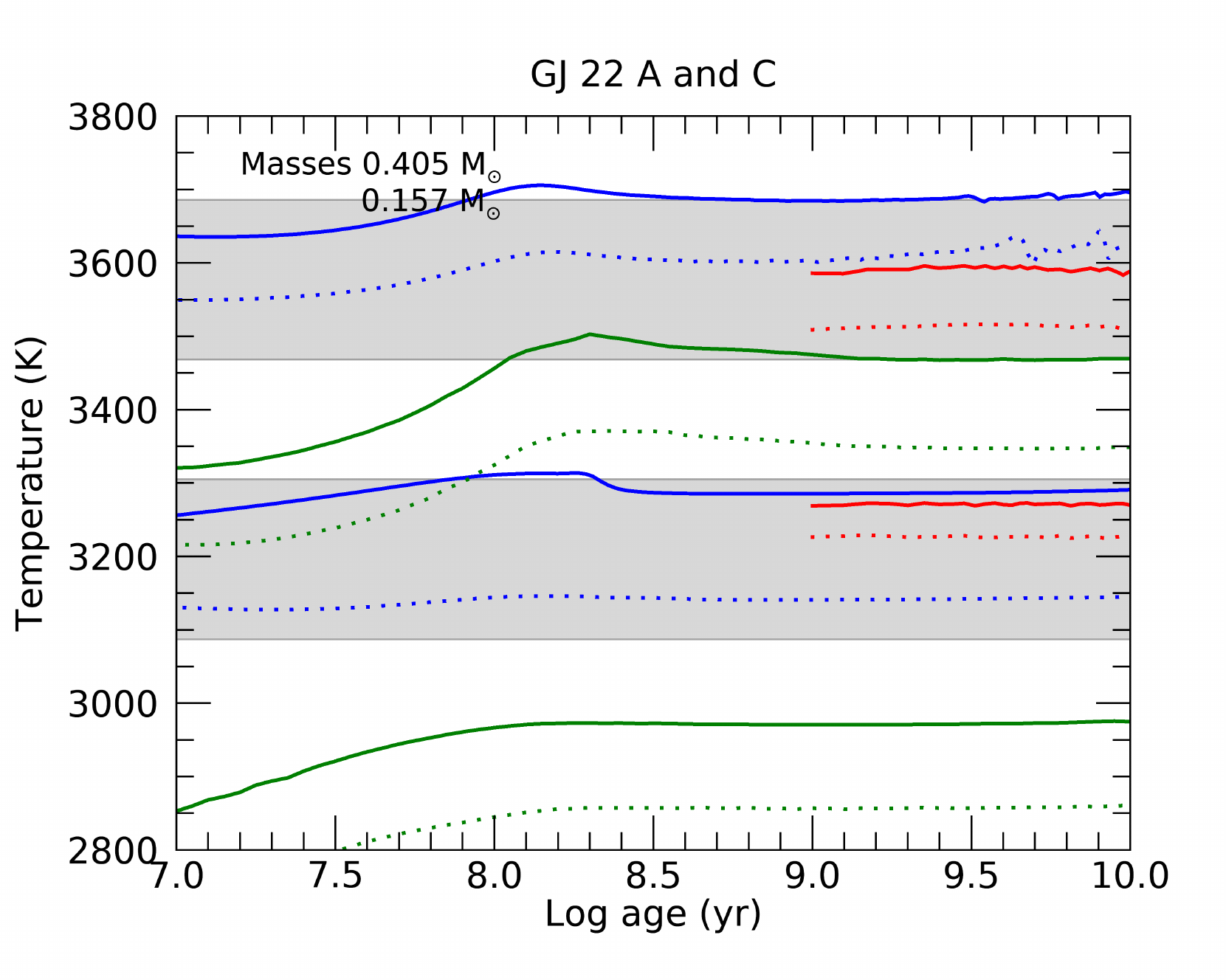}{0.39\textwidth}{}
           }
  \gridline{\fig{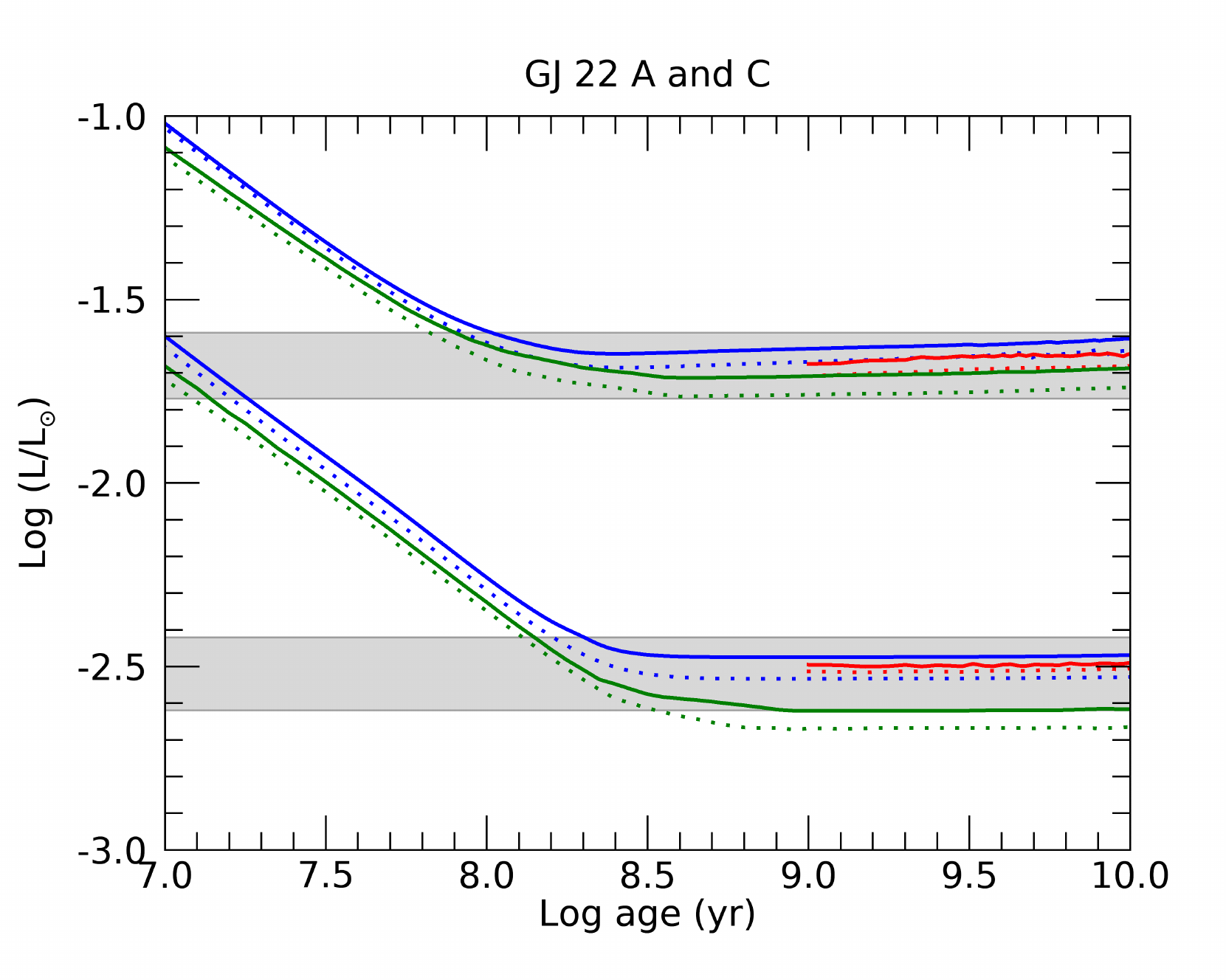}{0.39\textwidth}{}
           }
  \gridline{\fig{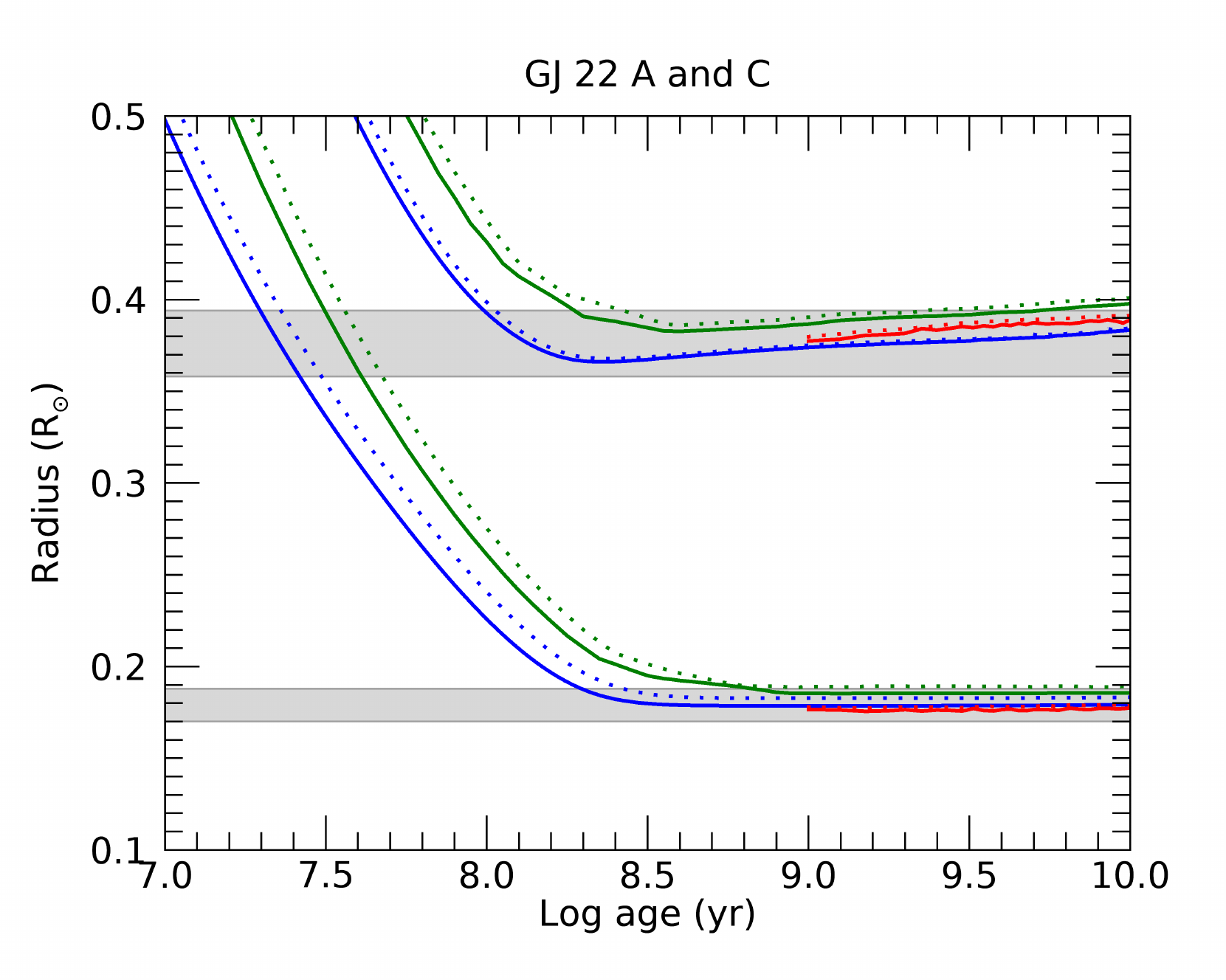}{0.39\textwidth}{}
           }
  \caption{\scriptsize Evolutionary tracks for GJ 22 A and C for models that
    vary metallicity. The dotted lines indicate solar metallicity while
    the solid lines indicate $[Fe/H] = -0.2$ to approximate the metallicity
    of the GJ 22 system ($[Fe/H] = -0.19$). Blue lines represent the MESA/MIST
    models, red lines represent the Dartmouth models, and green lines
    show the PARSEC models. At main sequence ages radius remains nearly invariant
    while temperature and luminosity increase with decreasing metallicity. 
    The shaded areas show the uncertainties inferred from the atmospheric fits
    (Table \ref{tab:temp}). Both metallicities can be accommodated by the data.
    \label{fig:metallicity}}
\end{figure}

Out of the five star systems we used to test models the GJ 22 AC
system stands out as the only system for which the models of stellar
structure and evolution were able to produce an accurate and self
consistent solution.  This is true of all models except for the PARSEC
model, which has systematic problems with under predicting effective
temperatures (Section \ref{subsubsec:parsec}). As discussed in Section
\ref{subsubsec:atmconsiderations}, the GJ 22 system is known to be
slightly metal poor, with \citet{Rojas-Ayalaetal2012} finding $[Fe/H]
= -0.19$, in agreement with our model fit. The system can therefore be
used as a model for the effect of deviations in metallicity for stars
with known masses. Out of the five evolutionary models we consider in
this work three have fine enough metallicity grids to model the effect
of of a change in metallicity of -0.19 dex. These are the Dartmouth
models, the MESA/MIST models, and the PARSEC models (Sections
\ref{subsubsec:dartmouth}, \ref{subsubsec:mist}, and
\ref{subsubsec:parsec}, respectively). Figure \ref{fig:evol1} shows
these three models plotted with $[Fe/H] = -0.20$ while the models of
\citet{Baraffeetal2015} and the YaPSI models remain at solar
metallicity.  Figure \ref{fig:metallicity} shows the evolutionary
tracks of the three models that encompass lower metallicities plotted
at both solar metallicity and $[Fe/H] = -0.20$. For a given mass there
is an increase in temperature and a corresponding increase in
luminosity, while radius remains mostly unchanged for main sequence
ages. The variations due to this small change in metallicity appear to
be contained within the uncertainties of the atmospheric models. Not
counting the PARSEC models, Figures \ref{fig:evol1} and
\ref{fig:metallicity} show equally acceptable fits for both solar
metallicity models and the models with reduced metallicity.  The cause
of the good evolutionary fits to GJ 22 A and C therefore appears to
not be connected to any variation in metallicity, which could have
been indicative of problems with the solar zero points adopted by the
several different models.

Our data do not support any further explanation for the fact that the
models provide such a good match to the GJ 22 AC system. We note that
while the spectroscopic fits GJ 22 A and C are good, so are the ones
for other stars in the sample. GJ 22 C also exhibits strong H$\alpha$
emission, as is common for mid to late M dwarfs, so a lack of magnetic
activity cannot be invoked as a simplifying factor either.
Another possible explanation is that in joint light $V\,sin\,i < 4\,km/s$ for
GJ 22 AC, indicating that both components are slow rotators \citep{Reinersetal2012}.

We suggest that further comparative studies between the GJ 22 AC system
and other systems could be particularly instructive with regards to
what is and is not working in stellar models.

\subsection{G 250-29 B and GJ 469 B: The Effects of Small Changes in Mass and Metallicity} \label{subsec:smallchange}
\begin{center}
\begin{figure*}[h!]
   \includegraphics[scale=0.6]{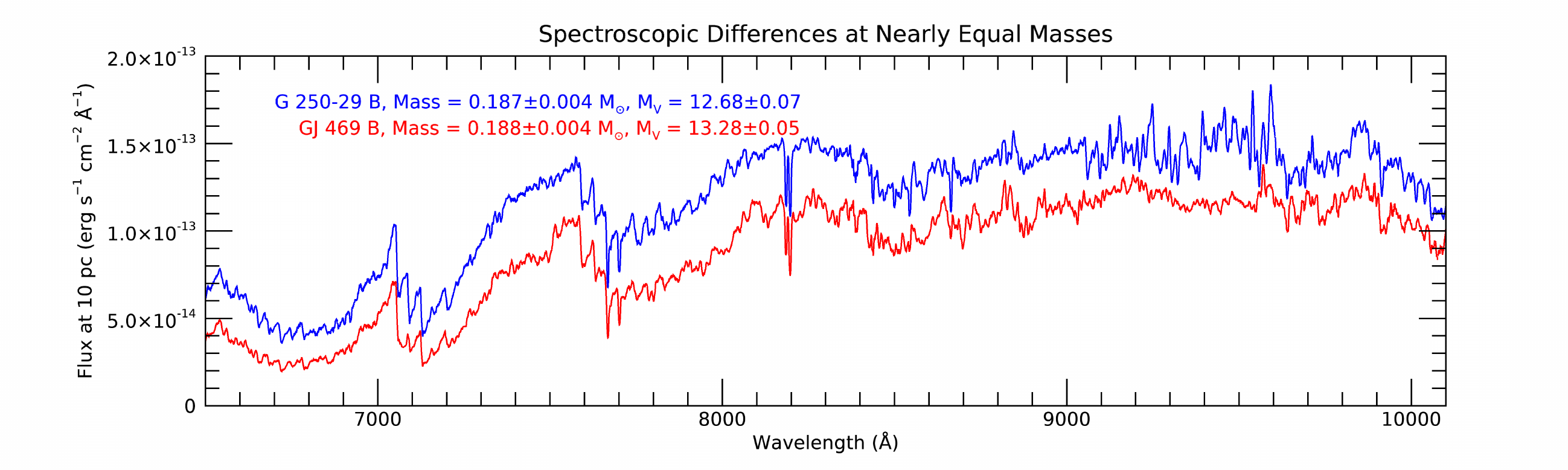}
   \caption{\scriptsize Flux calibrated spectra for G 250-29 B and GJ469 B.
     The spectra were smoothed using a Gaussian kernel for clarity. G 250-29 B
     has about 1.5 times the flux of GJ 469 B. Fringing is
     present at wavelengths greater than 9,000\,\AA.}
     \label{fig:specdifferences}
\end{figure*}
\end{center}

G 250-29 B and GJ 469 B provide an interesting example of how stars
with very similar masses and metallicities can vary significantly in
luminosity and temperature.  Figure \ref{fig:specdifferences} shows the
spectra for G 250-29 B (0.187$\pm$0.004\,M$_{\odot}$) and GJ 469 B
(0.188$\pm$0.004\,M$_{\odot}$). While the spectra are remarkably
similar in morphology the spectrum of G 250-29 B has about 1.5 times the
flux of GJ 469 B. Using the atmospheric derivations listed in Table
\ref{tab:temp}, G 250-29 B is more luminous by a factor of 1.26, still
within the uncertainties, and hotter by 145\,K, which is significant
given the uncertainty in temperature of 109\,K (Section
\ref{subsubsec:interferometry}). Their radii are also significantly
different at 0.231$\pm$0.011\,R$_{\odot}$ for G 250-29 B and
0.266$\pm$0.011\,R$_{\odot}$ for GJ 469 B.  Neither system shows signs of youth,
with no H$\alpha$ emission, calculated $Log\,g = 5.0$, and well fit Ca
and K gravity indicators. There may be a slight difference in
metallicity, with metallicities ([Fe/H]) of  -0.14 and -0.11 for the A and
B components of G 250-29 B, respectively, and -0.10 and -0.07 for the
A and B components of GJ 469. These differences are only borderline in
significance given that we can only infer the equal metallicities of components
of the same binaries to about 0.1 dex,  however they
do work in the conventional sense of making the most metal poor stars
hotter. In the case of the GJ 22 system we saw that a
significantly greater difference in metallicity of -0.19 had the effect
of changing the predicted model temperatures by only about 100\,K
(Figure \ref{fig:metallicity}), therefore either the models are
unreliable in their treatment of metallicity or it is unlikely that
such a small change in metallicity between G 250-29 B and GJ 469 B
would account for such a large change in observable characteristics.

This comparison between G 250-29 B and GJ 469 B shows that even with
very similar masses measured to high precision two stars can be
significantly different. The reasons for these differences are not
clear, and that adds a note of caution when interpreting M dwarf
evolutionary models. There are still higher order effects that
probably cannot be understood given our current constraints on
observational parameters and our ability to model them.

\subsection{The Transition to Full Convection, the Jao Gap,  and
  the Convective Kissing Instability} \label{subsec:kissing}

The transition to a fully convective interior is a hallmark of M
dwarfs. It is predicted to occur at masses ranging from
0.28\,M$_{\odot}$ to 0.33\,M$_{\odot}$
\citep[e.g.,][]{ChabrierAndBaraffe1997}, corresponding to early to mid
M subtypes. Since then several works have attempted to refine our
understanding of this transition.  Understanding this transition has
become particularly interesting in light of the so called {\it Jao
  gap} \citep{Jaoetal2018}, a thin gap in the color-magnitude diagram
noticed in {\it Gaia} DR2 data that is thought to be related to the
transition to full convection.

Theoretical work by \citet{vanSadersAndPinsonneault2012} propose that
there exists a mass range immediately above the onset of full
convection where $^3$He burning produces a convective core that is
initially separated from the star's deep convective zone by a thin
radiative envelope. As the convective core grows, periodic merging
with the convective envelope causes pulsations in luminosity,
temperature, and radius.  They call this phenomenon the {\it
  convective kissing instability}.  Their work uses the MESA stellar
evolution code (Section \ref{subsubsec:mist}) The MIST evolutionary
tracks plotted in Figures \ref{fig:evol1} and \ref{fig:evol2} show
those oscillations for three stars: G 250-29 A (0.35\,M$_{\odot}$), GJ
469 A (0.33\,M$_{\odot}$), and GJ 1081 A (0.32\,M$_{\odot}$).
\citet{BaraffeAndChabrier2018} also note the existence of the
convective kissing instability, but at a much narrower mass range of
0.34\,M$_{\odot}$ to 0.36\,M$_{\odot}$.  While that work predicts
pulsations, they do not appear in the \citet{Baraffeetal2015}
evolutionary tracks for G 250-29 A (0.35\,M$_{\odot}$, Figure
\ref{fig:evol2}) because that work produced a model grid in steps of
0.1\,M$_{\odot}$ as opposed to 0.01\,M$_{\odot}$ in
\citet{BaraffeAndChabrier2018}. Our interpolation therefore skipped
over this feature.  \citet{MacDonaldAndGizis2018} also postulate that
the Jao gap is caused by the increase in luminosity due to the merging
of a convective core and a convective envelope originally separated by
a thin radiative zone, but do not find that this merging leads to a
periodic instability.  As discussed extensively in
\citet{Jaoetal2018}, the YaPSI models (Section \ref{subsubsec:yapsi})
predict the Jao gap even though we do not see any manifestation of
pulsations in the YaPSI plots in this work. Without finer mass
coverage it is impossible to generalize the discussion to other
models, and a general assessment of issues regarding the convective
kissing instability or other features that may be causing the Jao gap
is not our goal. Our intent here is only to note an interesting
feature that we saw in a model set (the MIST models) in light of a
recent discovery (the Jao gap), and provide some context.

As seen in the above discussion, several theoretical issues with
observational implications arise in the mass range bordering the
transition from partial to full convection. It would be interesting to
test whether or not the convective kissing instability exists by
detecting a relation in fundamental parameters that follows the
pulsations predicted by the MIST models for  G 250-29 A, GJ
469 A, and GJ 1081 A. Similarly, the steeper slope of the mass luminosity
relation around the transition to full convection predicted by the YaPSI
models should be
tested observationally.  While the dynamical masses in our data set
are precise enough, we lack the very large sample with finely spaced
mass coverage that would be required for such tests. We therefore
emphasize that even with the robust mass-luminosity relation of
\citet{Benedictetal2016} there are still open questions in low mass
stellar structure whose answers will require the study of many more
systems with dynamical masses.

\begin{deluxetable*}{rccccc}
\tabletypesize{\tiny}
\tablecaption{Model Matches \tablenotemark{a} \label{tab:matches}}
\tablehead{ \colhead{Star, Property}   &
            \colhead{MIST}             &
            \colhead{Dartmouth}        &
            \colhead{Bar. 2015}        &
            \colhead{PARSEC}           &
            \colhead{YaPSI}            }
\startdata     
GJ 22 A $T_{eff}$   & \checkmark \checkmark \checkmark  & \checkmark \checkmark \checkmark  & \checkmark \checkmark \checkmark  & X                      & \checkmark \checkmark \checkmark   \\
$L$                & \checkmark \checkmark \checkmark  & \checkmark \checkmark \checkmark  & \checkmark \checkmark \checkmark  & \checkmark Y           & \checkmark \checkmark \checkmark   \\
$R$                & \checkmark \checkmark \checkmark  & \checkmark \checkmark \checkmark  & \checkmark \checkmark \checkmark  &  \checkmark \checkmark & \checkmark \checkmark \checkmark   \\
\hline                                                                                                                                                   
GJ 22 C $T_{eff}$   & \checkmark \checkmark \checkmark  & \checkmark \checkmark \checkmark  & \checkmark \checkmark \checkmark  & X                      & \checkmark \checkmark \checkmark   \\
$L$                & \checkmark \checkmark \checkmark  & \checkmark \checkmark \checkmark  & \checkmark \checkmark \checkmark  & YY                     & \checkmark \checkmark \checkmark   \\
$R$                & \checkmark \checkmark \checkmark  & \checkmark \checkmark \checkmark  & \checkmark \checkmark \checkmark  & \checkmark \checkmark  & \checkmark \checkmark \checkmark   \\  
\hline
\hline
GJ 1081 A $T_{eff}$ & \checkmark \checkmark             & \checkmark  \checkmark            & \checkmark \checkmark             & X                      & \checkmark \checkmark   \\
$L$                & \checkmark \checkmark             & \checkmark \checkmark             & \checkmark \checkmark             & YY                     & \checkmark \checkmark   \\
$R$                & X                                 & X                                 & YY                                & \checkmark \checkmark  & \checkmark              \\
\hline                                                                                                                                                       
GJ 1081 B $T_{eff}$ & \checkmark \checkmark             & \checkmark \checkmark             & \checkmark \checkmark             & X                      & \checkmark \checkmark   \\
$L$                & \checkmark \checkmark             & \checkmark \checkmark             & \checkmark \checkmark             & YY                     & \checkmark \checkmark   \\
$R$                & X                                 & X                                 & YY                                & \checkmark \checkmark  & \checkmark              \\
\hline                                                  
\hline                                                    
G 250-29 A $T_{eff}$ & \checkmark \checkmark            & \checkmark \checkmark             &  \checkmark  \checkmark          & X                       & \checkmark \checkmark    \\ 
$L$                 & \checkmark Y                     & \checkmark                        &  \checkmark Y                    & YY                      & \checkmark \checkmark    \\
$R$                 & Y                                & \checkmark                        &  \checkmark Y                    & \checkmark Y            & \checkmark Y             \\
\hline                                                                                                                                                      
G 250-29 B $T_{eff}$ &  \checkmark \checkmark           & \checkmark \checkmark             &  \checkmark \checkmark           & X                       & \checkmark \checkmark    \\
$L$                & YY                                & X                                 & Y                                & YY                      & \checkmark \checkmark    \\
$R$                & Y                                 & X                                 & YY                               & Y                       & YY                       \\
\hline
\hline
GJ 469 A $T_{eff}$  & X                                 & \checkmark                        & X                                 & \checkmark              &  \checkmark \checkmark  \\  
$L$                & \checkmark \checkmark             & \checkmark \checkmark             & \checkmark Y                      & \checkmark Y            & \checkmark \checkmark   \\
$R$                & \checkmark Y                      & \checkmark Y                      & \checkmark Y                      & \checkmark \checkmark   & \checkmark Y            \\
\hline                                                                                                                                                      
GJ 469 B $T_{eff}$  & \checkmark                        & X                                 &  \checkmark                       & X                       &  \checkmark \checkmark  \\
$L$                & \checkmark \checkmark             & \checkmark \checkmark             & YY                                & YY                      & \checkmark \checkmark   \\
$R$                & Y                                 & YY                                & YY                                &  \checkmark \checkmark  & YY                      \\
\hline
\hline                        
GJ 1245 A $T_{eff}$ & \checkmark                        & X                                 & \checkmark                       & X                        & \nodata     \\ 
$L$                & \checkmark                        & X                                 & \checkmark \checkmark            & \checkmark               & \nodata     \\
$R$                & \checkmark                        & \checkmark                        & \checkmark                       & X                        & \nodata     \\
\hline                                                                                                                           
GJ 1245 C $T_{eff}$ & \nodata                           & \nodata                           & Y                                & X                        & \nodata     \\
$L$                & \nodata                           & \nodata                           & \checkmark \checkmark            & \checkmark \checkmark    & \nodata     \\
$R$                & \nodata                           & \nodata                           & X                                & X                        & \nodata     \\
\hline
\hline
\enddata
\tablenotetext{a}{ \checkmark \checkmark \checkmark $-$ The model is a full match for the system. All parameter are correct and  mutually respect coevality.
  This condition is only satisfied for the MIST model of  GJ 22 AC. \\
  \checkmark \checkmark $-$ The parameter in question is predicted correctly for main sequence ages and respects coevality between the two components of the system,
  but is not coeval with the other parameter predictions for the same system. \\
  \checkmark Y $-$ The parameter in question is predicted correctly for both main sequence and pre main sequence ages, but coevality is only satisfied at pre
  main sequence ages. \\
  \checkmark $-$ The parameter in question is predicted correctly for main sequence ages, but the coevality condition between components
  of the same system is either not met or cannot be established. \\
  YY $-$ The parameter is only predicted correctly if the system is pre main sequence and in that case coevality is respected. \\
  Y $-$ The parameter is only predicted correctly if the star system is pre-main sequence. Coevality is not established.
  This condition is easy to satisfy due to the shape of most evolutionary tracks in Figures (\ref{fig:evol1}, \ref{fig:evol2}, and \ref{fig:evol3}), and is
  not necessarily indicative of a young system. \\
  X $-$ The parameter is not predicted correctly under any assumption. }
 \end{deluxetable*}
 \begin{deluxetable*}{rccccc}
\tabletypesize{\tiny}
\tablecaption{Radius Comparisons  \tablenotemark{a} \label{tab:radius}}
\tablehead{ \colhead{Star, Property}   &
            \colhead{MIST}             &
            \colhead{Dartmouth}        &
            \colhead{Bar. 2015}        &
            \colhead{PARSEC}           &
            \colhead{Yapsi}            }
\startdata     
GJ 22 A     & \checkmark   & \checkmark   & \checkmark   & \checkmark   & \checkmark   \\
C           & \checkmark   & \checkmark   & \checkmark   & \checkmark   & \checkmark   \\
GJ 1081 A   & $\Uparrow$   & $\Uparrow$   & $\Uparrow$   & $\Uparrow$   & $\Uparrow$   \\
B           & $\Uparrow$   & $\Uparrow$   & $\Uparrow$   & \checkmark   & $\Uparrow$   \\
G 259-29 A  & $\Uparrow$   & \checkmark   & $\Uparrow$   & \checkmark   & \checkmark   \\
B           & $\Uparrow$   & $\Uparrow$   & $\Uparrow$   & $\Uparrow$   & $\Uparrow$   \\
GJ 469 A    & \checkmark   & \checkmark   & \checkmark   & \checkmark   & \checkmark   \\
B           & $\Uparrow$   & $\Uparrow$   & $\Uparrow$   & \checkmark   & $\Uparrow$   \\
GJ 1245 A   & \checkmark   & \checkmark   & \checkmark   & $\Downarrow$ & \nodata      \\
C           & \nodata      & \nodata      & $\Downarrow$ & $\Downarrow$ & \nodata      \\
\enddata
\tablenotetext{a}{\checkmark means the model radius matches the radius we infer. \\
  $\Uparrow$ means the radius is inflated in the sense that theory predicts a smaller radius. \\
  $\Downarrow$ means the theoretically predicted radius is larger than what we infer.}
 \end{deluxetable*}

\subsection{Other Models} \label{subsec:approaches}
 In the current study we test the hypothesis that 
evolutionary models can produce model grids applicable to a
wide range of stellar masses, and obtained mixed results. One
limitation of the grid approach is that it becomes difficult to
treat second order effects such as rotation and magnetism.  It is
particularly noteworthy than none of the models discussed in Section
\ref{subsec:evolutionary} include magnetism.

It is not our goal here to judge the merits of models we did not
include in our tests, however it is worth noting that other approaches
to stellar modeling exist, and that many of them attempt to model the
effects of magnetism, rotation, and other higher order factors.
Significant work has been done in extending conventional models into
the magnetic domain with the incorporation of magnetohydrodynamics,
with emphasis on its effects on convection and radius inflation
\citep[e.g.,][]{FeidenAndChaboyer2012, MullanAndMacDonald2001}.  These
models are usually tested on a small number of stars with well known
properties.  Examples of low mass stars for which these models were
applied are: KOI-126 \citep{Feidenetal2011,SpadaAndDemarque2012}, EF
Aquarii \citep{FeidenAndChaboyer2012}, UV PSc, YY Gem, and CU Cnc
\citep{FeidenAndChaboyer2013}, Kepler-16 and CM Dra
\citep{FeidenAndChaboyer2014}, UScoCTIO5 and HIP 78977
\citep{Feiden2016}, LSPM J1314+1320 \citep{MacDonaldAndMullan2017a},
LP 661-13, KELT J041621-620046, and AD 3814
\citep{MacDonaldAndMullan2017b}, GJ 65 A \citep{MacDonaldetal2018},
and Trappist-1 \citep{Mullanetal2018}.

Unfortunately few of these tests used stars with well measured dynamical masses,
so the fundamental connection between mass and stellar evolution is often
tested only indirectly. Due to the work we present here the field is now ripe
for a new generation of model testing when theorists can use this data set
to fine tune model predictions.

\section{Conclusions and Future Work}\label{sec:conclusions}

The results from our tests are mixed. On one perspective, it is clear
that, with the exception of GJ 22 AC, the models cannot provide fully
consistent solutions to the full extent of the binary star evolution
test. From another perspective, the conditions of this test are quite
stringent, and we should not dismiss the fact that the models do have
predictive power.  It is also important to keep in mind that the tests
must be interpreted on a statistical context because while models
themselves are theoretical and do not carry uncertainties, their
comparisons to data do.  The data we are testing against, namely the
matches between observed spectra and model atmospheres summarized in
Table \ref{tab:temp} are matches that as an ensemble carry an
uncertainty, quoted and propagated to 1\,$\sigma$. Because those
atmospheric model comparisons are the root for evolutionary model
comparisons and they have been validated to 1\,$\sigma$ to the quoted
uncertainties we should expect the comparisons, or predictions, made
by the evolutionary models to also be correct only two thirds of the
time, regardless of the fact that models themselves are not
constructed to a certain level of uncertainty.  We should further take
into account that, despite the significant amount of observing
resources used by this project, the test sample remains small. A
sample of ten stars in five binary systems is prone to uncertainties
arising from statistics of small numbers.  Nevertheless, being in full
agreement with the data only in one out of five systems is likely to
be a deficit beyond statistical uncertainties.  As previously noted,
agreement between models and observations could be reached if the
uncertainties in the quantities derived from atmospheric models were
artificially increased. In that sense it is clear that atmospheric
models are further along in predictive power than evolutionary models
when it comes to predicting the same basic stellar parameters.  On the
other hand, analysis of the similarities and differences between G
250-29 B and GJ 469 B (Section \ref{subsec:smallchange}) indicates
that the problem of M dwarf modeling may be intrinsically more complex
than what we imagine, with stars of similar masses and metallicities
having significantly different observable parameters. If that is a
general case then it could be that the expectation we have for model
results are simply not realistic.

We believe that the best way to interpret the results we present here is to say
that evolutionary models should be used with caution. In an age when
the drive to characterize exoplanets places a large emphasis
on stellar parameters, the accuracy of parameters derived from
evolutionary models should not yet be taken for granted, as they often are.

We believe, however, that the true value of the
data we present here is
its potential to test models that are specifically built or fine-tuned
to the dynamical masses and spectra of each binary component, while
respecting the constraints of coevality and equal metallicity natural
to a binary system. As such, we see these observations and the present
work not only as a means to test the past,
but rather as a tool to guide future theoretical efforts. We note that
all spectra discussed here are available as a digital supplement to
this work and we encourage theorists to use them as a means of
constraining new models.

On the observational front we note the need of similar observations to
extend the mass coverage to the late M dwarfs, where our analysis was
based on only one binary system, GJ 1245 AC. We note also that while
broad mass coverage is valuable, detailed observations of systems of
nearly equal mass with well known dynamical masses are essential to
make sense of secondary effects such as rotation and magnetic field
topology, especially around the transition to full convection.  We
plan to carry out similar observations for such systems in the near
future.

\section{Summary}\label{sec:summary}
We used HST/STIS to obtain spatially resolved spectra of five M dwarf
systems with known individual dynamical masses and used them as
benchmarks to test models of stellar structure and evolution. Our
principal findings are as follows.
\begin{itemize}

\item{The BT-Settl atmospheric models produce synthetic spectra that
  are a good match to observations, and their validity was verified
  by comparison to parameters derived with long baseline optical interferometry
  (Section \ref{subsubsec:interferometry}). We adopt their best
  match temperature as an approximation of the true effective
  temperature of our targets. The agreement is somewhat worse at
  cooler temperatures, possibly due to the need for a finer
  temperature grid and also the intrinsic complexity in modeling
  cooler atmospheres due to molecules and dust formation
  (Section \ref{subsubsec:atmconsiderations}).}

\item{There may be a weak tendency for the BT-Settl models to underestimate
  surface gravities (Section \ref{subsubsec:atmconsiderations}).}

\item{We tested the Dartmouth evolutionary models
  \citep{Dotteretal2008}, the MIST evolutionary models
  \citep{Choietal2016,Dotteretal2016}, the models of
  \citet{Baraffeetal2015}, the PARSEC models  \citep{Bressanetal2012}, and
  the YaPSI models \citep{Spadaetal2017}
  with the properties derived from our
  comparison of observed spectra to model atmospheres. We find only
  marginal agreement between evolutionary models and observations.
  Out of five systems the models only reproduced one of them, GJ 22 AC,
    in a self consistent manner. We note that the PARSEC models are systematically
    too cold (Section \ref{subsec:evolutionary}, Figures \ref{fig:evol1},
  \ref{fig:evol2}, and \ref{fig:evol3}, and Table \ref{tab:matches}).}

\item{We  confirm the known
  tendency towards radius inflation, in the sense that models
  under-predict the true radius (Section \ref{subsec:radius}).}
  
\item{We note that the GJ 22 AC system is well modeled in a self-consistent manner
  by all models we tested except for the systematically too cold PARSEC models.
  The system is slightly metal poor, but that does not seem to affect the quality
  of the fits. It is not clear what if anything is special about the GJ 22 system
  (Section \ref{subsec:gj22ac}).}

\item{We discuss the case of G 250-29 B and GJ 469 B, where nearly equal masses and metallicities
  produce significantly different luminosities, temperature and radii.
  This example may be indicative of the need for a more detailed treatment of stellar
  structure and evolution (Section \ref{subsec:smallchange}).}
  
\item{We note that the principal utility of the data presented here is
  not as a test of existing models but rather as a guide for future
  theoretical approaches. As such, we include all data as a digital
  supplement (Section \ref{sec:conclusions}).}

\item{We emphasize the need for more spatially resolved spectroscopic observations of
  M dwarfs with dynamical masses, especially for masses close to the transition to
  full convection (Section \ref{subsec:kissing} and for late M dwarfs, where our coverage
  consists of a single system, GJ 1245 AC (Section \ref{sec:conclusions}).}
\end{itemize}

\acknowledgments{
We thank Russel White, Alycia Weinberger, G. Fritz Benedict, and the
STIS instrument team at STScI for helpful
discussions. We are grateful to Andrew Mann for providing his calibration spectra.
S. B. D. acknowledges support from the NSF Astronomy and
Astrophysics Postdoctoral Fellowship program through grant
AST-1400680. T. J. H. acknowledges support from NSF grant AST-171555.
Support for HST-GO program number 12938 was provided by
NASA through a grant from the Space Telescope Science Institute, which
is operated by the Association of Universities for Research in
Astronomy, Incorporated, under NASA contract NAS5-26555.
}

\bibliography{/Users/sergiodieterich/sergesreferences2}

\end{document}